\documentclass[seceq,supplement]{ptptex}
\usepackage{graphicx}

\usepackage{epstopdf}
\usepackage{color}

\def\al{\alpha}  
\def\be{\beta} 
\def\ga{\gamma}
\def\de{\delta}
\def\ep{\epsilon}

\def\et{\eta}
\def\th{\theta}

\def\ka{\kappa}
\def\la{\lambda}

\def\si{\sigma}
\def\ta{\tau}

\def\om{\omega}
\def\De{\Delta}
\def\Ga{\Gamma}

\def\La{\Lambda}

\def\Om{\Omega}

\def\bk{{\mathbf{k}}}
\def\bn{{\mathbf{n}}}

\def\bx{{\mathbf{x}}}

\def\bX{{\mathbf{X}}}

\def\D{{\cal D}}

\newcommand{\ben}{\begin{equation}}
\newcommand{\een}{\end{equation}}
\newcommand{\bea}{\begin{eqnarray}}
\newcommand{\eea}{\end{eqnarray}}
\newcommand{\ba}{\begin{array}}
\newcommand{\ea}{\end{array}}
\newcommand{\bit}{\begin{itemize}}
\newcommand{\eit}{\end{itemize}}

\newcommand{\secthead}[1]{%
	 \subsection{#1}
}

\def\pa{\partial}

\def\half{\frac{1}{2}}

\newcommand{\vev}[1]{\left\langle#1\right\rangle}

\def\dprime{\mathaccent"707D}
\def\vb{\bar v}
\def\bdi#1{\ensuremath{\boldsymbol{#1}}}
\def\bx{{\bdi x}}

\def\bX{{\bdi X}}

\def\bk{{\bdi k}}

\def\Xd{\dot X}

\def\Xp{\acute X}

\def\bXd{\dot\bX}
\def\bXp{\acute\bX}
\def\bXpp{\dprime{\bX}}

\def\tb{\bar{t}}

\newcommand{\Obhh}{\ensuremath{\Omega_{\rm b}h^{2}}}

\newcommand{\fd}{\ensuremath{f_{\rm 10}}}

\newcommand{\mpl}{\ensuremath{m_\mathrm{P}}}

\definecolor{lightblue}{rgb}{0.85,0.9,1}
\definecolor{lightgreen}{rgb}{0.8,1,0.8}
\definecolor{lightyellow}{rgb}{1,1,0.5}
\definecolor{darkgreen}{rgb}{0,0.6,0}
\definecolor{darkred}{rgb}{0.6,0,0}
\definecolor{darkblue}{rgb}{0,0,0.6}
\definecolor{grey}{rgb}{0.5,0.5,0.5}
\definecolor{lightgrey}{rgb}{0.2,0.2,0.2}

\newcommand{\ms}{\ensuremath M_{\rm s}}
\newcommand{\mcs}{\ensuremath M_{\rm cs}}
\newcommand{\horDis}{\ensuremath d_h}
\newcommand{\powspec}{\mathcal{P}}

\title{
Signals of Inflationary Models with Cosmic Strings%
}

\author{
Mark \textsc{Hindmarsh}%
}

\inst{
Department of Physics and Astronomy\\
University of Sussex\\
Brighton BN2 1QB\\
U.K.}

\abst{
A class of well-motivated models of inflation end by producing cosmic strings.  The current status
of efforts to calculate and observe the signals from such models are outlined, with a particular emphasis on cosmic strings, and on the Cosmic Microwave Background signal.
}

\begin{document}

\maketitle



\section{Inflationary cosmology and cosmic strings}

Current observations are consistent with a remarkably simple model of the very early universe: 
inflation \cite{Inflation}.
At a time of about $10^{-36}$ sec all the energy density of the universe is held in the form of a homogeneous scalar field, the inflaton.  The potential energy density relaxes so slowly that it acts like an effective cosmological constant, causing the expansion of the universe to accelerate.  In this accelerating or inflating phase, the quantum fluctuations in the scalar field are amplified, blown up in scale, and frozen in as a scale-invariant spectrum of density fluctuations 
\cite{PertsFromInf}.
The inflaton eventually approaches the minimum of its potential energy density, oscillates around it,  and decays into other particles, signalling the start of the conventional hot big bang. Thanks to the period of inflation the universe is large, flat, and smooth, apart from the scale-free perturbations laid in by the field.


In some models there is a phase transition 
in a companion set of fields towards the end of inflation \cite{Shafi:1984tt,HybInf}.  A thermal phase transition in a theory with a topologically non-trivial set of ground states leads to the formation of defects \cite{DefFor}, 
extended structures made from spatially varying configurations of the scalar (and possibly also gauge) fields.  
At the end of inflation the inflaton replaces the temperature as the control parameter, and defects form equally well 
\cite{Shafi:1984tt,DefForInf,Yokoyama:1988zza}. Hybrid models can be naturally accommodated in the framework on supersymmetry and Grand Unification \cite{Dvali:1994ms}.

Of particular interest are effectively one-dimensional structures, cosmic strings \cite{CosStrRev,recentCSreviews}, where the energy density is concentrated into a thin tube.  Strings have the crucial property: they decay just fast enough to maintain a constant density relative to the rest of the contents of the universe, a property known as scaling. This property is not shared by all defects: in particular monopoles (point-like) and domain walls (two-dimensional) end up dominating the energy density by today (13 GYr) unless their mass scale is much lower than that of inflation ($10^{15}$ GeV). 

One-dimensional structures are also singled out as the fundamental constituents of matter in string theory, which allows for infinitely long and relatively light (compared with the Planck scale) ``elementary" strings to be created in the hot big bang \cite{FDstringForm}.
These strings also scale, and hence the study of cosmic strings is now an important feature of the cosmology of elementary strings.



Although the interactions of strings are model-dependent, all kinds of strings have gravitational fields, and if they are massive enough they add to the gravitational perturbations from inflation. The search for strings is therefore partly a search for deviations from the standard inflationary model.
This work focuses on the signals from strings, as the predictions of inflationary cosmology are now part of the standard lore \cite{Lyth:2009zz}, with a particular emphasis on the Cosmic Microwave Background (CMB) signal where there is an interesting interplay between the inflation and cosmic string parameters. It also concentrates on ``ordinary" cosmic strings: those without currents, without junctions, and with unit reconnection probability, to remain to a certain extent complementary to other recent reviews \cite{recentCSreviews}.


\section{Cosmic strings}

Cosmic strings are linear distributions of mass-energy in the universe, which may have any length.    They are characterised by two numbers: the mass per unit length $\mu$, and their tension $T$. In their fundamental form, the tension and the mass per unit length are equal, $\mu = T$, which means that there is a boost invariance along the length of the string and they are relativistic.

Like particles, they may be elementary or composite -- made from more fundamental constituents, in this case scalar and possibly gauge fields. If they are made from fields, they are classical solutions of the field equations localised into tubes of energy density, related to solitons \cite{Manton:2004tk}. Classically, elementary strings have zero width, and solitonic strings have a non-zero width set by a mass scale in the classical field equations which we denote $\mcs$.  Quantum fluctuations smear out an elementary string to a size which is the inverse string scale $\ms^{-1}$, but do not significantly correct the size of a solitonic string in a weakly coupled field theory \cite{Preskill:1986kp}.

%
%
%
In both cases, the classical dynamics are simple for strings whose curvature radius is much larger than the width (either $\mcs^{-1}$ or $\ms^{-1}$):  the acceleration is proportional to the local curvature, 
with certain constraints.  Strings couple to other fields; certainly to gravity, with a strength $G\mu$, where $G$ is Newton's constant, and $\mu \sim \mcs^2$ or $\ms^2$.
For solitonic strings whose mass per unit length is set by the Grand Unification scale (about $10^{15}$ GeV), $G\mu$ is in the range $10^{-7}$ -- $10^{-6}$, while for elementary strings the effective four-dimensional tension depends on the details of the compactification of the extra dimensions.  In the benchmark KKLMMT warped compactification scenario \cite{Kachru:2003sx}, which includes a mechanism of inflation, phenomenological considerations put the string tension in the range $10^{-10} \lesssim G\mu \lesssim 10^{-6}$ \cite{Firouzjahi:2005dh}, with a similar range for unwarped compactifications \cite{FDstringForm}.

Strings also couple to massive states: solitonic strings are constructed from the fields of particles with mass $\mcs$, and elementary string excitations come in an infinite tower of states with masses $M^2 \sim n\ms^2$, where $n$ is an integer.  One can also regard string loops as massive states, produced by self-intersections of the string.  

\begin{figure}
\includegraphics[width=0.5\textwidth]{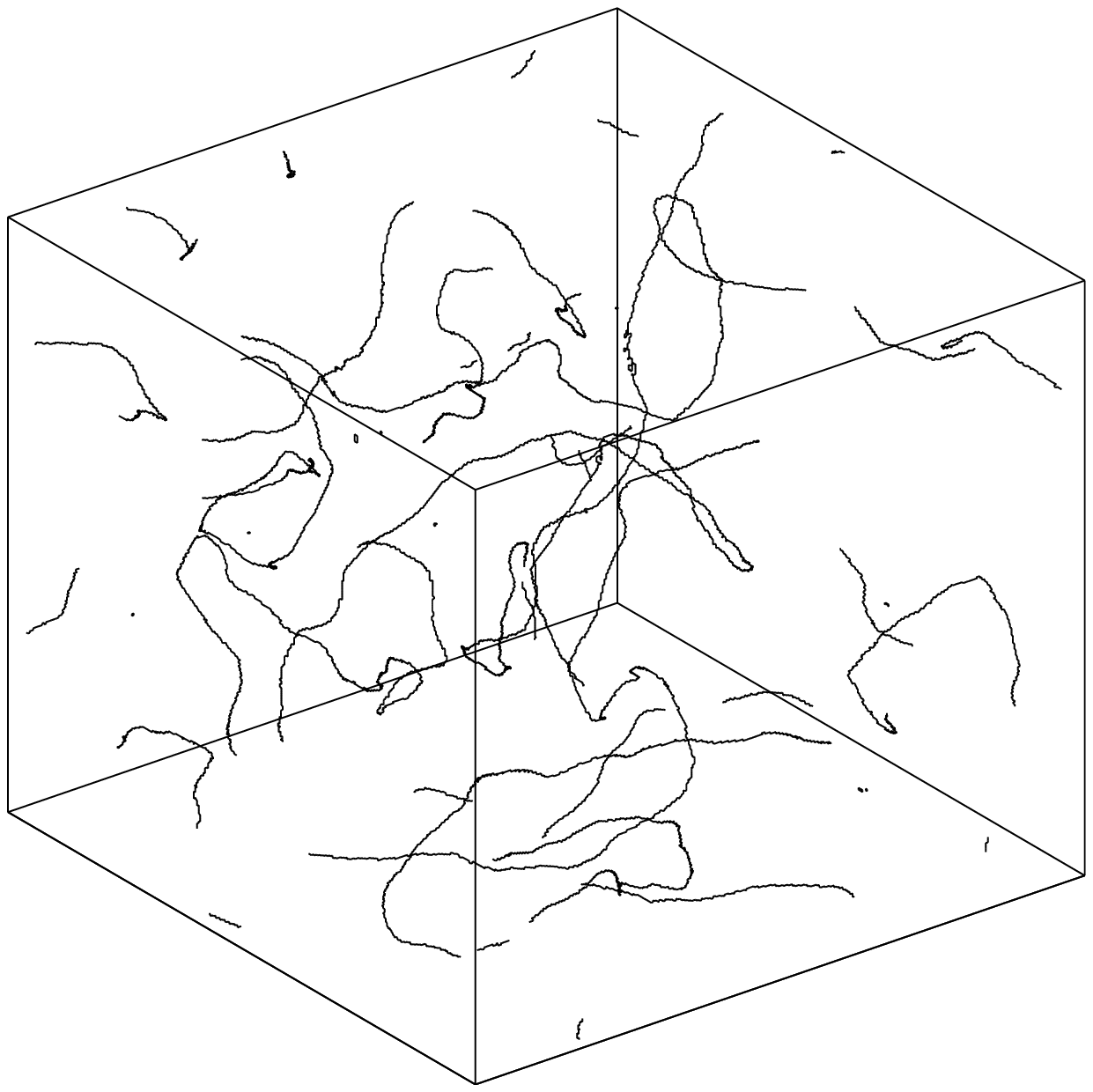}
%
\includegraphics[width=0.5\textwidth]{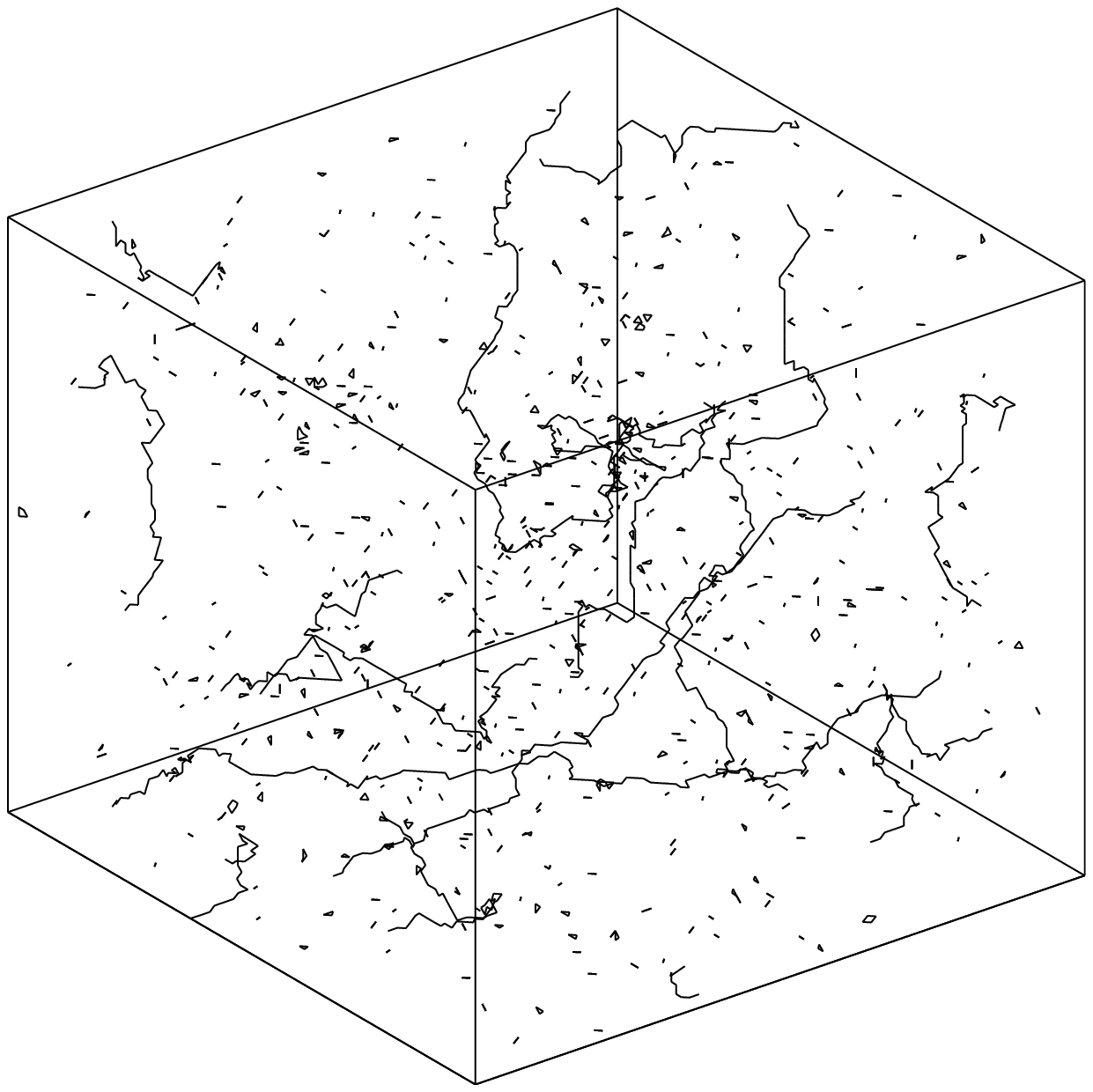}
\caption{\label{f:StringBoxes} String networks from simulations performed using in classical field theory (left) and in the Nambu-Goto approximation (right). Each box represents an approximately horizon-sized volume at any epoch in the history of the network.  Note that the ``infinite" strings, stretching across the box, are rather similar in curvature radius and density, but that loops are almost entirely absent in the field theory simulation.}
\end{figure}

An unsolved problem in the cosmic string scenario is the eventual destination of the energy in the string network.  It has been addressed in different ways: originally by numerical simulations of classical relativistic (Nambu-Goto) strings 
\cite{OrigNGsim,OluVan,BlancoPillado:2011dq}
and then by simulations of a classical field theory with string solutions 
\cite{Vincent:1997cx,Moore:2001px,Hindmarsh:2008dw}. 
The similarities and  differences between the two approaches is easily seen in Figure \ref{f:StringBoxes}, which shows a horizon-sized volume from an example of each.  Thanks to the scaling property, the snapshots could be taken at any time. The strings are divided between ``infinite" strings, stretching across the box, and loops.  The infinite strings rather similar in curvature radius and length density, but loops are almost entirely absent in the field theory simulation.  The reason for this is that all the strings' energy density goes into massive modes of the fields, which do not show up in the figure.  In the Nambu-Goto approximation, this channel does not exist, and instead the energy of the infinite strings goes into small loops.  
These loops are assumed to decay slowly into gravitational radiation.

It should be noted in neither set of simulations has gravitational radiation been included, which is vital in the traditional string scenario to determining the average size of the loops when they are produced: gravitational radiation reaction on the infinite strings is supposed to set this scale. In the scenario based on the classical field theory, it is argued that the gravitational fields are O($G\mu$), and therefore the gravitational radiation reaction is negligible compared with the reaction from the massive modes, which has an O(1) effect on the string energy density over a Hubble time 
\cite{Hindmarsh:2008dw}.

\section{Observational signals from strings}

There are many ways to search for strings in the universe, but as
outlined above there 
is a source of uncertainty in the predictions of the scenario. 
Perhaps the most well-explored avenue exploits the gravitational fields of the strings, which perturb the matter and radiation around them.  The amplitude of these perturbations is controlled by the dimensionless parameter 
$8\pi G\mu$, where $G$ is Newton's constant and $\mu$ is the string mass per unit length. The factor of $8\pi$ comes from Einstein's equations. Hence gravitational perturbations of amplitude $10^{-6}$ -- $10^{-5}$ are to be expected from strings in Grand Unified Theory, right in a very interesting observational range.

The best-developed calculations are those for the power spectrum of the Cosmic Microwave Background fluctuations.  Strings generated the fluctuations both by stirring up the photon-baryon fluid in the surface of last scattering, and by perturbing the CMB photons on their journey towards us 
\cite{GKS, Bouchet:1988hh}. 
Standard codes for computing the CMB temperature and polarisation power spectra 
\cite{EBsolvers} are easily adapted to accommodate sources of energy-momentum, and there is good understanding of how the sources depend on length scale and time, derived from the combination of 3D numerical simulations with modelling at various levels. The results from numerical simulations 
\cite{Allen:1996wi,Contaldi:1998mx,Landriau:2003xf,Bevis:2006mj,Bevis:2010gj,Landriau:2010cb}
depend somewhat on the modelling, but the differences can be simply
parametrised in a phenomenological model of string evolution.
\cite{USM,Wyman:2005tu,Battye:2006pk,Pogosian:2007gi,Pogosian:2008am,Battye:2010xz}
They are often quoted in terms of $\fd$, the fraction of the power due to strings at a multipole $\ell = 10$, and depending on the modelling and the dataset the upper limit 95\% confidence limit upper bound is currently in the range 2 -- 10\% 
\cite{Sievers:2009ah,Battye:2010xz,Dunkley:2010ge}.

The gravitational fields of the strings likewise perturb the matter in the universe, although strings are much more efficient at making CMB perturbations than matter density fluctuations. This is because strings are a source of vector and tensor gravitational perturbations as well as scalar ones: all three produce CMB fluctuations, but only scalar perturbations affect the density.  The same codes which compute the CMB power spectrum can also supply the power spectrum of linear density perturbations from strings, showing that a string component consistent with the CMB is negligible except at very small scales 
\cite{Wyman:2005tu}. 
String-induced matter perturbations are rather different from those sourced by inflation, as they are non-linear from the outset: matter is drawn in behind a moving string into a thin sheet or wake 
\cite{Wakes}. 
This means that standard linear perturbation theory is inaccurate. The dominance of the string perturbations at very small scales means that the galaxy formation may proceed very differently even with a small component of strings (as measured by the CMB power). In particular the bias (the square root of ratio of the galaxy power spectrum to the matter power spectrum) is subject to uncertainty, and therefore directly applying the galaxy linear power spectrum 
\cite{Tegmark:2006az}
to constrain the matter power spectrum in models with strings is not reliable.  Constraints derived from observations of the Baryon Acoustic Oscillation 
\cite{Percival:2009xn} 
in large-scale structure surveys are not so sensitive to the bias, and therefore should be applicable to models with strings.

The gravitational fields of strings can also produce gravitational lensing of galaxies, quasars, and other distant compact sources 
\cite{Vilenkin:1984ea,Hindmarsh:1989yn,deLaix:1996vc,Sazhin:2006kf,Mack:2007ae}. 
A gravitational lens produced by a straight string consists of a pair of unamplified and undistorted images, quite different from those produced by galaxies and clusters of galaxies, and it would seem that lens searches are an excellent way to look for strings.  However, string lenses must be separated from the background signal of pairs of similar galaxies \cite{Hindmarsh:1989yn,Morganson:2009yk}. Furthermore, the maximum separation of the images is approximately $5 (G\mu/10^{-6})$ arc sec, which is to be compared with the resolution of ground-based surveys and the angular size of galaxies at redshifts of $z \sim 1$, both of which are around 1 arc sec. Future radio instruments such as the Square Kilometer Array \cite{Koopmans:2004gf} could produce interesting constraints from surveys of Compact Radio Objects \cite{Mack:2007ae}, perhaps even as low as $G\mu \lesssim 10^{-9}$.




The gravitational signal with the largest theoretical uncertainty is gravitational radiation 
\cite{Vilenkin:1981bx,Sakellariadou:1990ne,Hindmarsh:1990xi,Allen:1991bk,Caldwell:1991jj,Damour:2000wa,Siemens:2006yp}. 
Gravitational radiation is emitted both from infinite strings, from the collision of small amplitude waves 
and from oscillating string loops.  The power from a string loop of any size is \cite{Vilenkin:1981bx} $P = \Ga G\mu^2$, where $\Ga$ is a O(100) shape-dependent factor: the total power per unit volume from a string network is therefore proportional to the number density of loops.  This number density depends very strongly on assumptions made about the principal energy loss mechanism for the network.  Classical field theory simulations show that there is O(1) loop per horizon volume  surviving for one Hubble time \cite{Hindmarsh:2008dw}, while the traditional cosmic string scenario argues that the number per horizon volume is much larger, $(\Ga G\mu)^{-3p}$ in a universe whose scale factor depends on time as $a \propto t^p$.  The root of this difference is that in classical field theory simulations, strings can decay into the massive particles of the field from which they are constructed, whereas the traditional scenario assumes that this channel is eventually closed by the strings becoming smooth, and that gravitational radiation takes over. 

Similar remarks apply to other forms of massless radiation: both solitonic and fundamental strings may couple to massless pseudoscalar fields, the prime example being axion strings \cite{Vilenkin:1982ks}.  Massless pseudoscalar radiation is also emitted by oscillating strings \cite{Vilenkin:1986ku}, although there is disagreement over the power and the spectrum \cite{Battye:1993jv,Hagmann:2000ja}.  Classical field theory simulations again show O(1) loop per horizon volume 
\cite{Yamaguchi:1998gx}.

Solitonic strings in a realistic theory incorporating the Standard Model must couple at some level to known particles, and so should be a source of cosmic rays, perhaps even Ultra-High Energy cosmic rays (UHECRs) with energies above $10^{10}$ GeV 
\cite{Bhattacharjee:1998qc}
if their mass scale is high enough.  Again, the power in cosmic rays is uncertain, but there are essentially two scenarios: classical field theory simulations tell us that all the energy of the string network goes into massive particles, and therefore there is a potentially large fraction of the total available energy going into extremely energetic Standard Model particles 
\cite{Vincent:1996rb,Vincent:1997cx},
which can show up both as UHECRs and, via a cascade of decays and interactions, as GeV-scale $\ga$-rays. 
From cosmic ray observations there are strong but model-dependent constraints on the energy injection rate and hence on the mass density in cosmic strings 
\cite{Sigl:1995kk,Protheroe:1996pd,Bhattacharjee:1997in,Sigl:1998vz,Wichoski:1998kh}. 
In the traditional scenario, particles are produced only near cusps (rare regions of the string moving close to the speed of light) and consequently the constraints are much weaker. The decay products may still be of interest for producing baryon asymmetry through out-of-equilibrium decays 
\cite{CSBaryogenesis}
or as an extra source of dark matter beyond standard thermal freeze-out \cite{Jeannerot:1999yn}.

\section{Cosmic strings from high-energy physics}

\subsection{Solitonic string solutions in the Abelian Higgs model}
\label{ss:NOstring}

The simplest gauge field theory showing a string solution is the Abelian Higgs model \cite{Nielsen:1973cs}.  It consists of a complex scalar $\phi$ and a gauge field $A_\mu$, with a Minkowski space action
\ben
S = \int d^4x\, \left( D_\mu\phi^*D^\mu\phi +V(\phi) + \frac{1}{4}F^{\mu\nu}F_{\mu\nu}\right),
\een
where the covariant derivative is $D_\mu = \partial_\mu - ieA_\mu$ and the potential energy density is $V = \half \la \left( |\phi|^2 - \phi_0^2\right)^2$.
The field equations are 
\bea
D^2\phi + \la\left( |\phi|^2 - \phi_0^2\right)\phi &=& 0 \\
\pa^\mu F_{\mu\nu}  - ie(\phi^*D_\nu\phi - D_\nu\phi^*\phi) &=& 0.
\eea
The ground state can be written as
$\phi = e^{i\al}\phi_0$, $A_\mu = 0$, or gauge transformations thereof.  There are excitations around the ground state, which correspond to a scalar of mass $m_{\rm s} = \sqrt{2\la}\phi_0$ and a vector of mass $m_{\rm v} = e\phi_0$

\begin{figure}
\centering
\includegraphics[width=0.6\textwidth]{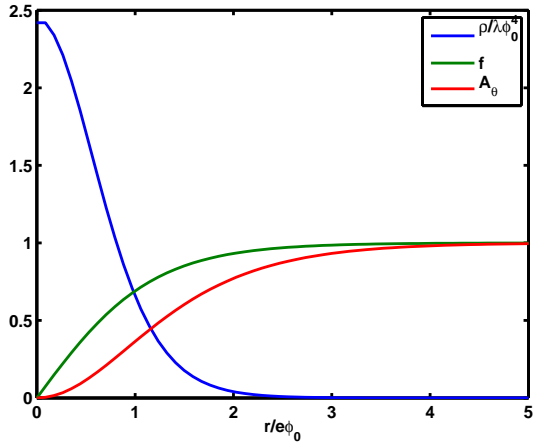}
\caption{\label{NOprofiles} Numerical solutions for the Nielson-Olesen vortex, for parameters $\la = e = 1$. Shown is the dimensionless energy density $\rho/(e\phi_0^2)$, the magnitude of the scalar field $f$, and the azimuthal component of the gauge field $A_\theta$, plotted as a function of the distance from the origin in units of the vector mass $e \phi_0$. }
\end{figure}

There exist cylindrically symmetric finite energy static solutions of the following form
\ben
\phi = \phi_0 f(r)e^{i\theta}, \quad A_i = \frac{1}{er}A_\th(r)\hat\th_i, \quad A_0 = 0,
\een
where $\th$ is an azimuthal coordinate and $r$ a radial one.  By substituting the ansatz into the field equations, one can discover numerical solutions of the form shown in Fig.\ \ref{NOprofiles}.  

One can see that the the energy density is confined to a region whose size is of order the inverse vector mass, $(e\phi_0)^{-1}$, and that it contains a tube of magnetic flux oriented in the $z$ direction, $B_z = A_\th'(r)/er$.  The string mass per unit length is found to be 
\ben
\mu = 2\pi B(\la/e^2) \phi_0^2, 
\een
where $B$ is a slowly varying function of its argument, approximately given by
\cite{Hill:1987qx}
\ben
B(\beta) \simeq \left\{ \ba{cc}1.04\beta^{0.195}, &  10^{-2} < \be \ll 1, \cr 
                                                   2.4/\ln(2/\be), &  \be < 10^{-2} .
                                                   \ea \right.
\een

\subsection{Inflation models with strings}
\label{ss:InfStrings}

The inflation scenario supposes that the energy density of the universe was dominated by a homogeneous gauge singlet scalar field $s$ (see e.g.\ Ref.\ \citen{Lyth:2009zz}) with potential $V(s)$. The key observational quantities are the power spectra of the metric perturbations, broken down into scalar and tensor components $\powspec_{s,t}(k)$, and their tilts $n_{s,t} = d \ln \powspec_{s,t}/d \ln k$.  In standard  slow-roll inflation driven by a single field,
\bea
\powspec_s(k) & \simeq & \left. \frac{1}{24\pi^2}\frac{V}{m_p^4}\frac{1}{\ep}\right|_{aH=k}, \quad
\powspec_t(k) \simeq \left. \frac{1}{6\pi^2}\frac{V}{m_p^4}\right|_{aH=k}, \\
n_s & \simeq & \left.(1 - 6\ep + 2\et)\right|_{aH=k}, \quad n_t = -2\ep,
\eea
where $\ep = \mpl^2 V''/(2V)$, $\eta = \mpl^2 V''/V$ are the slow-roll parameters, which must be small during inflation.
All quantities are evaluated at ``horizon crossing" $aH=k$, when the wavelength of the mode in question has grown to the Hubble length. 

In hybrid inflation models the inflaton can be coupled to the string scalar field \cite{Yokoyama:1988zza,HybInf}
\ben
V(s,\phi) = V_s(s) + \half\la(|\phi|^2 - \phi_0^2)^2 + h s^2 |\phi|^2.
\een
For large $s$, the local minimum of $\phi$ is at $\phi=0$, and the potential energy density is $V_s(s) + \half\la\phi_0^4$.  Inflation can take place provided $V_s$ is sufficiently flat, and terminates at a critical value $ s_c^2 = \la\phi_0^2/h$, where the $\phi$ field becomes unstable, rolls towards its minimum (where $V=0$) and the string-forming phase transition occurs.

When looking for flat potentials, it is natural to explore supersymmetric theories. A supersymmetric hybrid inflation model with a U(1) gauge symmetry can be constructed from three chiral superfields $S$, $\Phi$ and $\bar\Phi$, with U(1) charges $0$, $1$ and $-1$ respectively,  and superpotential \cite{Dvali:1994ms}
\ben
\label{e:superpot}
W = \ka S (\Phi\bar\Phi - M^2).
\een
Adopting lower case letters for the scalar components, the scalar potential is then
\ben
V(s,\phi,\bar\phi) = \ka^2|s|^2(|\phi|^2 + |\bar\phi|^2) + \ka^2|\phi\bar\phi - M^2|^2 + \frac{g^2}{2}(\xi -|\phi|^2 + |\bar\phi|^2)^2,
\een
where $g$ is the U(1) gauge coupling and a D-term with Fayet-Iliopoulos (FI) parameter $\xi$ has been added.  One can see that the charged fields become unstable at $|s_c|^2 = (M^2 + \xi g^2/\ka^2)$. If the FI term dominates the potential energy density, we have D-term inflation \cite{DtermInf}, and if subdominant (or absent) F-term inflation (see e.g. \citen{Lyth:1998xn}).

The classical potential is completely flat at $\phi = \bar\phi = 0$, with $V = \ka^2M^4 + \half g^2\xi^2$, and one needs to take into account the radiative corrections for the system to be able to evolve towards its minimum.  For $|s| \gg |s_c|$ one has 
\ben \Delta V_s = 
\frac{1}{16\pi^2}(\ka^2 M^4 + g^2\xi^2)\ln \left(\frac{\ka^2s^2}{\mu^2}\right),
\label{Vphiphib}
\een
where $\mu$ is the renormalisation scale.  In models where the inflation fields $(S,\Phi,\bar\Phi)$ are coupled to the supersymmetric standard model fields (see e.g.\ Refs.\ \citen{Garbrecht:2006az}) the potential is somewhat more complicated. 

This class of simple supersymmetric inflation models is highly predictive, as the cosmic string and inflationary scalar perturbation amplitude are to first order controlled by the same parameter $M^2 + \xi g^2/\ka^2$ \cite{Jeannerot:1997is}.  One is pushed towards the F-term model with a very small coupling $\ka$ \cite{Rocher:2004uv,Battye:2006pk}, or towards a non-minimal supergravity model \cite{Jeannerot:2005mc,Battye:2006pk}.

\secthead{Elementary cosmic strings}

Strings may be of any length, and so the question arises whether we could expect to find ``elementary" strings of the size of the cosmological horizon, $L \sim t$.  The question was considered by Witten \cite{Witten:1985fp}. Firstly, one can expect to find only the strings of a closed string theory: long open strings would fragment with a time of order the inverse string scale, $\ms^{-1}$.  Secondly, the strings will be very massive: Newton's constant is related to the string tension $1/(2\pi\al')$ by $G \sim \al'$, and so one expects $G\mu = \textrm{O}(1)$.  Thirdly, cosmic superstrings  would also be axionic, and therefore bounded by domain walls of surface density  $\La_{\rm QCD}^3$, where $\La_{\rm QCD} \sim 1$ GeV is the QCD scale.

Following the discovery of D-branes \cite{Polchinski:1995mt}, large extra dimensions \cite{ArkaniHamed:1998rs}, and warped compactification \cite{Randall:1999ee} the question was reconsidered. 
Importantly,  the new compactifications could reduce the effective 4-dimensional string tension well below the Planck scale.  For example, in a warped compactification scenario the metric takes the form
\ben
ds^2 = e^{2A(y)}\eta_{\mu\nu}dx^\mu dx^\nu + dy_6^2,
\een
where $y^a\;(a=1\ldots 6)$  are the compactified coordinates and $x^\mu\;(\mu=0\ldots 3)$ are coordinates in order 4-dimensional spacetime.  This means that a fundamental string living at $y = y_0$, where the ``warp factor" is $A_0$, has a tension
\ben
\mu = \frac{e^{2A_0}}{2\pi\al'}, 
\een
and with sufficient warping the string tension can be well below the Planck scale.  

Strings from string theory come in two kinds: the fundamental or F-strings of the original theory, and D-strings, which are 1-dimensional D-branes, objects on which fundamental open strings end (see e.g. \citen{Polchinski:1998rq})\footnote{To distinguish these strings from solitonic strings F-strings and D-strings are together called elementary in this review.}.  They can form bound states 
\cite{Schwarz:1995dk}
of $p$ F-strings and $q$ D-strings, with the composite $(p,q)$-string having tension in a flat 10-dimensional spacetime 
\ben
\bar\mu_{p,q} = \frac{1}{2\pi\al'}\sqrt{p^2 + g_s^{-2}q^2},
\een
where $g_s$ is the string coupling, which is of order unity. More complicated formulae apply in the presence of background fluxes and in warped compactifications.\cite{Firouzjahi:2005dh}
Provided these strings are well-separated in the extra dimensions from other, higher-dimensional, D-branes on which they can break, they are essentially stable. 

String theory therefore contains an interesting spectrum of objects which could constitute cosmic strings, and it also has a possible formation mechanism through the collision of D3 and anti-D3 branes at the end of brane inflation 
\cite{FDstringForm}. 
It turns out that this model looks rather like a  hybrid inflation model, with the inflaton being the brane separation modulus, and the waterfall field the tachyon fields associated with the string states connecting the two branes.


The low-energy effective action of a macroscopic quantum string is just the classical Nambu-Goto action with the string coupled to massless fields: perturbatively, these are the spacetime metric, the dilaton, and a Ramond-Ramond scalar (equivalently a 2-form field). 
On phenomenological grounds, the scalars should gain a potential as part of the compactification process, and as a result become massive and fixed in value.  Therefore we expect the strings with coordinate $X^\mu(\si^\al)$, $\al = 0,1$, moving in a 4-dimensional spacetime with metric $g^{(4)}$,  to be described by the action
\ben
\label{eq:NGaction}
S_{\rm s} = -  \int d^4 x \left( \frac{\mpl^2}{2}\sqrt{-g^{(4)}} R + \mu \int d^2\si \sqrt{-\ga} \de^{(4)}\left(x - X(\si)\right) \right),
\een
where $\ga _{\al\be} = g^{(4)}_{\mu\nu}\pa_\al X^\mu \pa_\be X^\nu$ is the induced metric on the string worldsheet, and $R$ is the 4-dimensional Ricci scalar.

An important feature of elementary cosmic strings 
is the presence of junctions.  A $(p,q)$-string can split into e.g.\ a $(p-1,q)$-string and an F-string at a three-way junction, a feature which has resulted in a lot of work on the classical dynamics of cosmic strings with junctions 
\cite{Junctions}.

Certain field theories, including the Nielson-Olesen string at $\la/e^2 < 1$, can also produce bound states and junctions, which means that the junction is not a distinguishing feature of an elementary cosmic string.  On the other hand, these theories can be used to study the dynamics of networks of strings with junctions \cite{NetJunct}.



%
%
%
%
%
%
%
%
%
%
%
%
%
%
%

\section{Dynamics of cosmic strings}

\subsection{Abelian Higgs model in an expanding background}
\label{ss:AHstrings}

In Section \ref{ss:NOstring} we exhibited the Nielsen-Olesen string solution of the Abelian Higgs model, a static and cylindrically symmetric field configuration representing a straight string.  One can expect to find more general time-dependent solutions representing moving strings, either infinite or in the form of closed loops.  If we wish to study these time-dependent solutions in a cosmological context, one must take the action for the field theory in a Friedmann-Robertson-Walker (FRW) background metric,
\[
S = -\int d^4x \, \sqrt{-g} \left( g^{\mu\nu}D_\mu\phi^*D_\nu\phi +V(\phi) + \frac{1}{4e^2}g^{\mu\rho}g^{\nu\si}F_{\mu\nu}F_{\rho\si}\right),
\]
where we have rescaled the gauge field $A_\mu \to A_\mu/e$.
The background metric is 
\ben
ds^2 = a^2(\tau)(-d\tau^2 + d\bx^2),
\een
where $d\bx^2$ is the 3-dimensional line element, and $a(\tau) \propto \tau$ or $ \tau^2 $ depending on whether the universe is radiation or matter-dominated.

In the temporal gauge ($A_0 = 0$)  the field equations are
\bea
\ddot\phi + 2\frac{\dot a }{a}\dot \phi -D^2\phi + \lambda a^2(|\phi|^2 -\phi_0^2)\phi &=& 0, \\
\pa^\mu\left(\frac{1}{e^2} F_{\mu\nu} \right) - ia^2(\phi^*D_\nu\phi - D_\nu\phi^*\phi) &=& 0,
\eea
where the index on the partial derivative is raised with the Minkowski metric.

One can solve this system of partial differential equations numerically. 
\cite{Vincent:1997cx,Moore:2001px,Bevis:2006mj}
The scalar field is placed on a cubic grid with lattice spacing $\De x$, with the gauge field on the links. The boundary conditions are toroidal, and the lattice spacing must be chosen so that the core of the strings can be resolved: in practice $\De x = 0.5 m_{\rm v}^{-1}$ is sufficient.
 
The initial conditions model an initial thermal or quantum vacuum state in which correlations are short-ranged: the simplest way to do this is to select the real and imaginary components of the scalar field from a gaussian distribution of zero mean and unit variance, independently at each site, at initial time $\tau_\textrm{i} = \De x$.  The subsequent evolution generates a random gauge field, with the Hubble damping term removing the high frequency modes and cooling the system so that the scalar field settles into the minima of the potential at $|\phi| = \phi_0$.  

There is a technical problem to solve:  the physical width of the string is fixed by the mass scale $m_{\rm v}$, and so the comoving width of string shrinks. as $a^{-1}$. Eventually the string will no longer be resolved by the lattice and becomes stuck on the grid.  One can adjust the lattice spacing so that $\De x = 0.5 m_{\rm v}^{-1}$ at the end of the simulation, however, this means that the width of the string is large at the initial time and the simulation box can contain fewer strings.  This is a practical strategy in the radiation era, where $a \propto \tau$, but not in the matter era, where $a \propto \tau^2$.

To get around the problem, one can modify field equations \cite{Ryden:1989vj,Bevis:2006mj}
according to 
\bea
\ddot\phi + 2\frac{\dot a }{a}\dot \phi -D^2\phi + \lambda a^{2s}(|\phi|^2 -\phi_0^2)\phi &=& 0,\nonumber \\
\pa^\mu\left(\frac{a^{2(1-s)}}{e^2} F_{\mu\nu} \right) - ia^2(\phi^*D_\nu\phi - D_\nu\phi^*\phi) &=& 0,\nonumber
\eea
If $s < 1$ the comoving width of the string does not shrink so fast, but the physical width of string \(w = m_{\rm v}^{-1}\) ``fattens'' as $a^{1-s}$.  The string tension remains constant as the ratio of the effective couplings is constant, which means that the collective dynamics of the strings is unaffected. 

This particular choice of modification preserves Gauss's Law (and hence current conservation) but because of the time-dependence violates covariant energy-momentum conservation.  
One must therefore carefully check the dependence of results on the adjustable parameter \(s\), which as mentioned above can be taken to the physical value \(s=1\) in the radiation era.


\begin{figure}

\centering

\includegraphics[width = 0.49\textwidth]{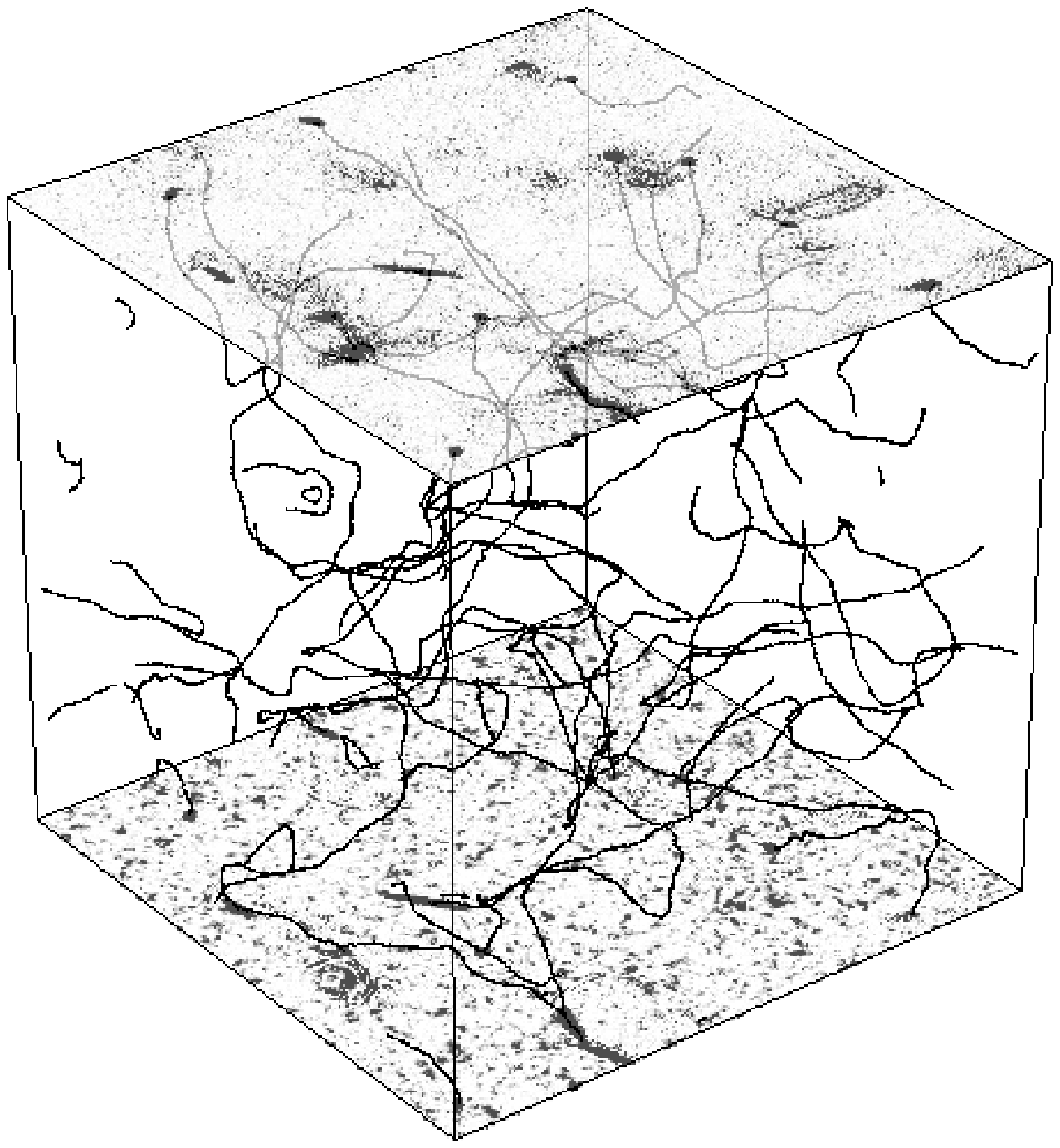}
\raisebox{10mm}{\includegraphics[width=0.49\textwidth]{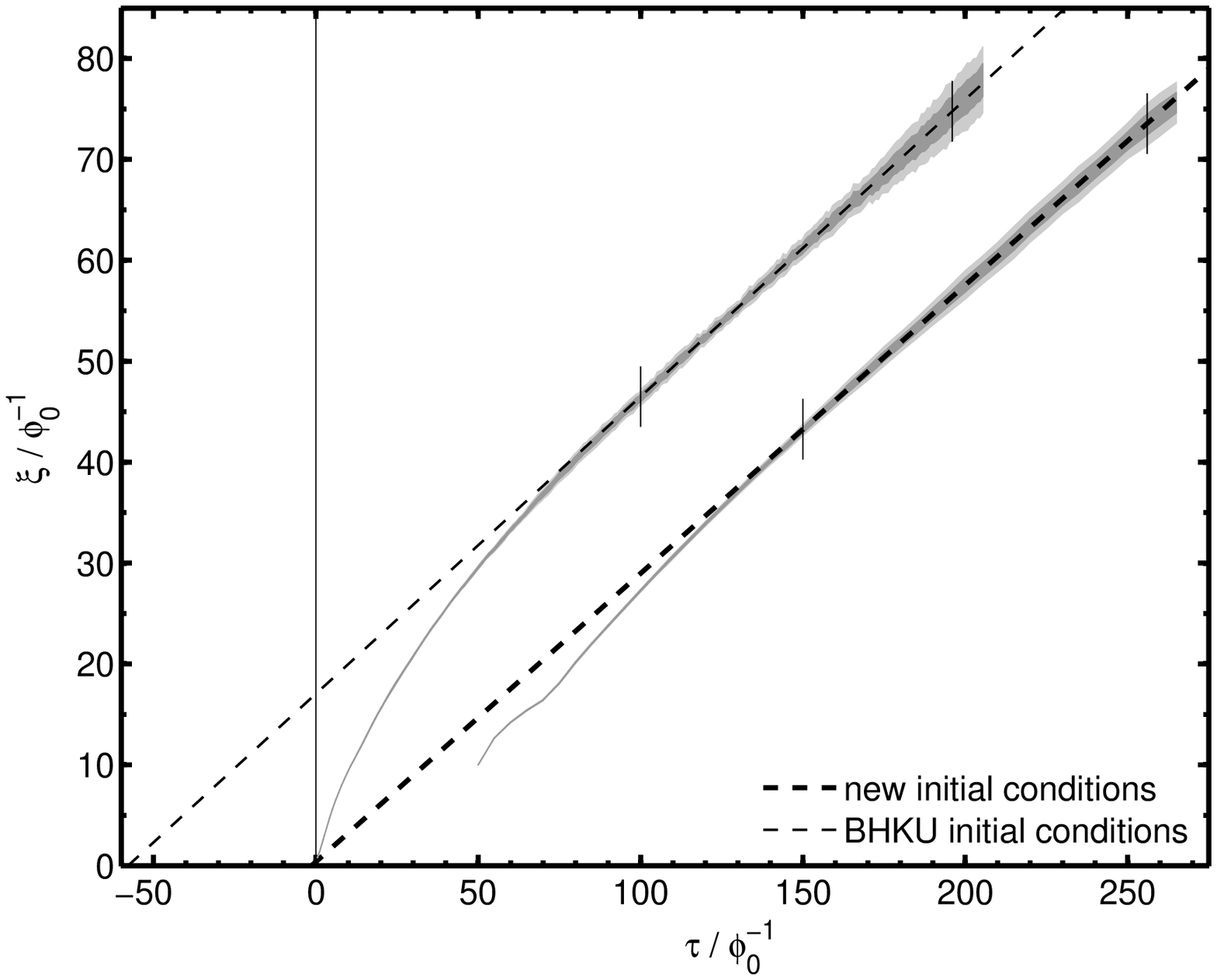}}

\caption{\label{f:AHsimPic}  Left: snapshot from a numerical simulation of a string network in the Abelian Higgs model, as described in the text.  The lines show the centres of the strings, and the shading on the top and bottom faces represents magnetic field energy density and scalar field potential
energy density respectively. Right: 
\label{f:xiCombine} 
string network length scale $\xi$ as a function of conformal time, averaged over two sets of 5 simulations with different initial conditions (see Ref. \citen{Bevis:2010gj} for an explanation).  After an initial relaxation period the length scale evolves linearly, meaning that the network is scaling. The parameters of the simulation were: $\la = 1$, $e = 1$, $\De x = 0.5$ at the end of the simulation, $\De t = 0.1$, volume $768^3$ (upper line) and $1024^3$ (lower line), with $s = 0.3,0$ respectively.
}  

\end{figure}

A snapshot from a simulation on a $512^3$ lattice in the radiation era is shown in Figure \ref{f:AHsimPic}. The random initial conditions on the field have produced a random network of strings, most of which stretches across the box and so represents infinite strings. The shading on the top and bottom faces picks out the fluctuations in the gauge and scalar field energy densities, generated by the decays of the strings themselves.  As the system evolves, the total length of string reduces, and all the energy eventually ends up in the form of gauge and scalar radiation. 

To be more quantitative, one can examine how the (comoving) length $L$ of string evolves with time.  It is convenient to look at the quantity 
\ben
\label{eq:xiDef}
\xi  = \sqrt(V/L),
\een
where $V$ is the comoving simulation volume, which can be thought of as
the approximate distance between strings.  Examining Figure
\ref{f:xiCombine}, 
one sees that after an initial relaxation period $\xi$ settles down to 
a linear growth in conformal time
\ben
 \xi \propto \ta,
\een
with the same constant of proportionality for two different ways of setting up the initial conditions \cite{Bevis:2010gj}.
One can measure the length in a number of different ways, but they are all proportional to each other.

A linear growth of the comoving length scale with conformal time is easily seen to imply that the physical length scale $\xi_{\rm p} = a\xi$ grows linearly with cosmological time $t = \int a(\tau) d\tau$.  Hence the physical length per unit volume is decreasing as $t^{-2}$.  Given that the energy per unit length is $\mu$, the physical energy density in string goes as 
\ben
\rho_{\rm s} \sim \frac{\mu}{t^2}.
\een
Hence the string density parameter $\Om_{\rm s} = 8\pi G \rho_{\rm s}/3H^2$ is constant,  $\Om_{\rm s} \sim G\mu$, a property known as scaling.  

\subsection{Nambu-Goto strings}
\label{ss:NGstrings}

In the limit that the ratio of the string width $w$ to the curvature radius $R$ goes to zero, one can treat solitonic strings as being ideal relativistic line objects 
\cite{Forster:1974ga,Carter:1994ag,Arodz:1995dg,Anderson:1997ip},
in which case they should be described by the Nambu-Goto action coupled to gravity (\ref{eq:NGaction}), and also a massless pseudoscalar in the case of a global string \cite{Vilenkin:1986ku}.  Large elementary strings are  also described by this action \cite{Tseytlin:1990vf}.

\begin{figure}
\centering

\includegraphics[width=0.5\textwidth]{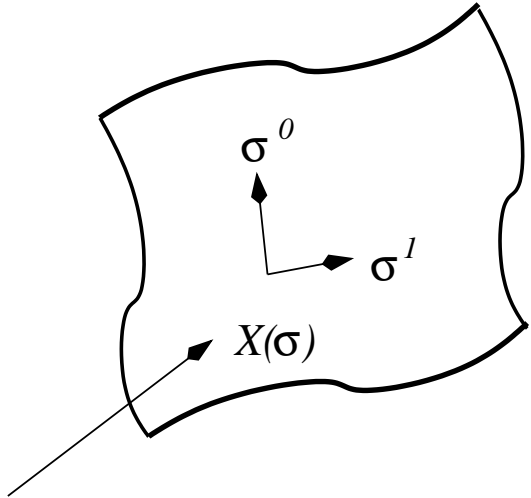}

\caption{\label{f:wsheet} Worldsheet of an ideal string in spacetime, labelled by coordinates $\si^0$ (timelike) and $\si^1$ (spacelike).}

\end{figure}

In studying the dynamics of Nambu-Goto strings it is convenient to chose worldsheet coordinates $\si^\al$ (see Fig.\ \ref{f:wsheet}) such that 
\ben
\Xd\cdot\Xp = 0, \quad \Xd^2 + \Xp^2 = 0,
\een
where the dot (prime) indicates a derivative with respect to $\si^0$ ($\si^1$), and  to choose the temporal or static gauge $X^0 = \si^0$, identifying worldsheet time with 4-d time. In this gauge the equations of motion in a FRW spacetime become
\bea
\ddot\bX + 2\frac{\dot a}{a}\left(1 - \bXd^2 \right) \bXd - \frac{1}{\ep} \frac{\pa}{\pa\si}\left(  \frac{1}{\ep}\frac{\pa \bX}{\pa\si}\right) &=& 0, \\
\dot\ep + 2 \frac{\dot a}{a}\left(1 - \bXd^2 \right) \ep &=& 0, 
\eea 
where
$
\ep = \sqrt{\bXp^2/(1-\bXd^2)}. 
$
Here, $\ep$ is the comoving energy per unit comoving length as measured by $\si \equiv \si^1$, and can be chosen arbitrarily at some initial time.

For short wavelength oscillations and on timescales much smaller than the Hubble time, the strings are effectively moving in Minkowski spacetime, for which we can choose $\ep = 1$, and the equations become
\bea
\ddot\bX - \bXpp = 0, \quad \bXd \cdot \bXp = 0, \qquad \bXd^2 + \bXp^2 = 1.
\eea
It is clear that the equations take the form of a wave equation, with unit propagation speed, supplemented by constraints which keep the velocity orthogonal to the tangent vector and the energy per unit length constant.

When strings cross, the classical Nambu-Goto action must be supplemented with a rule determined from the underlying theory.  Numerical simulations in the Abelian Higgs model 
\cite{Laguna:1989hn,Achucarro:2006es}
show that crossing strings reconnect, unless they are moving very close to the speed of light, or are nearly parallel. Network simulations show that this is rather unlikely, and that the reconnection probability averaged over all angles and collision speeds is close to 1. Elementary strings on the other hand, being quantum mechanical objects, reconnect with a probability 
\ben
P_{\rm r} \sim g_s^2 \frac{4\pi^2{\al'}^3}{V_\perp} f(\th,v),
\een 
where $g_s \sim 1$ is the string coupling, $f(\th,v)$ depends on the orientation $\th$ and relative speed $v$, and  $V_\perp$ is the volume of the extra dimensions in which they move 
\cite{Polchinski:1988cn}. 
This probability may be rather small, which is a key difference between elementary and solitonic strings: the lower the reconnection probability the higher the string length density.  Numerical simulations in flat space are consistent with $(L/V) \propto P_r^{-1}$  \cite{Sakellariadou:2004wq}, while in cosmological backgrounds $(L/V) \sim P_r^{-0.6}$ \cite{Avgoustidis:2005nv}.

\begin{figure}
\centering

\includegraphics[width=0.5\textwidth]{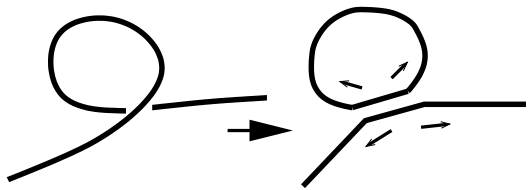}

\caption{\label{f:loopForm}  Reconnection of a string leading to loop formation.  Two pairs of oppositely-travelling kinks (discontinuities in the string tangent vector $\bX'$) are formed.}

\end{figure}

\begin{figure}
\centering
\includegraphics[width=0.5\textwidth]{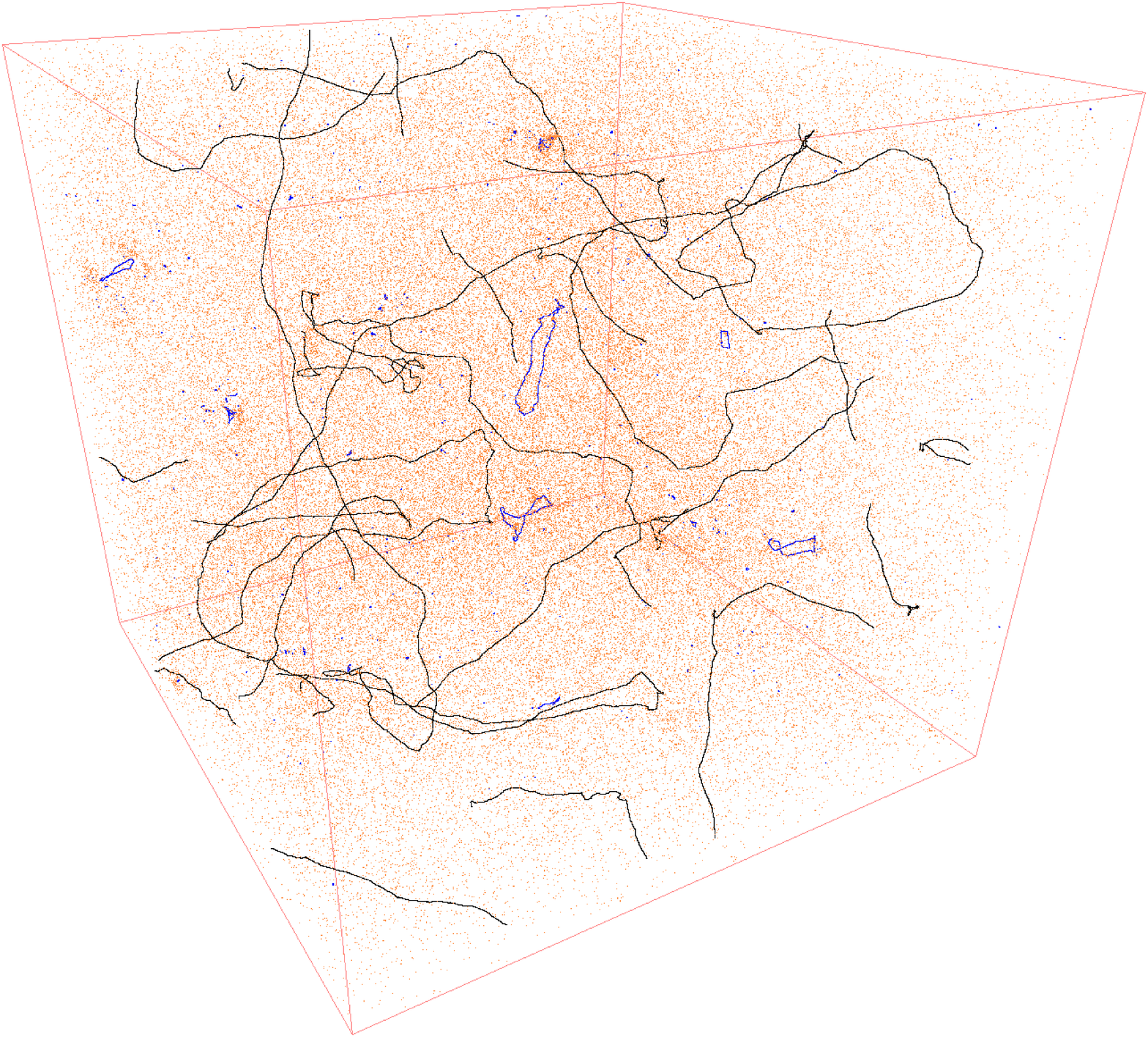}
\includegraphics[width=0.49\textwidth]{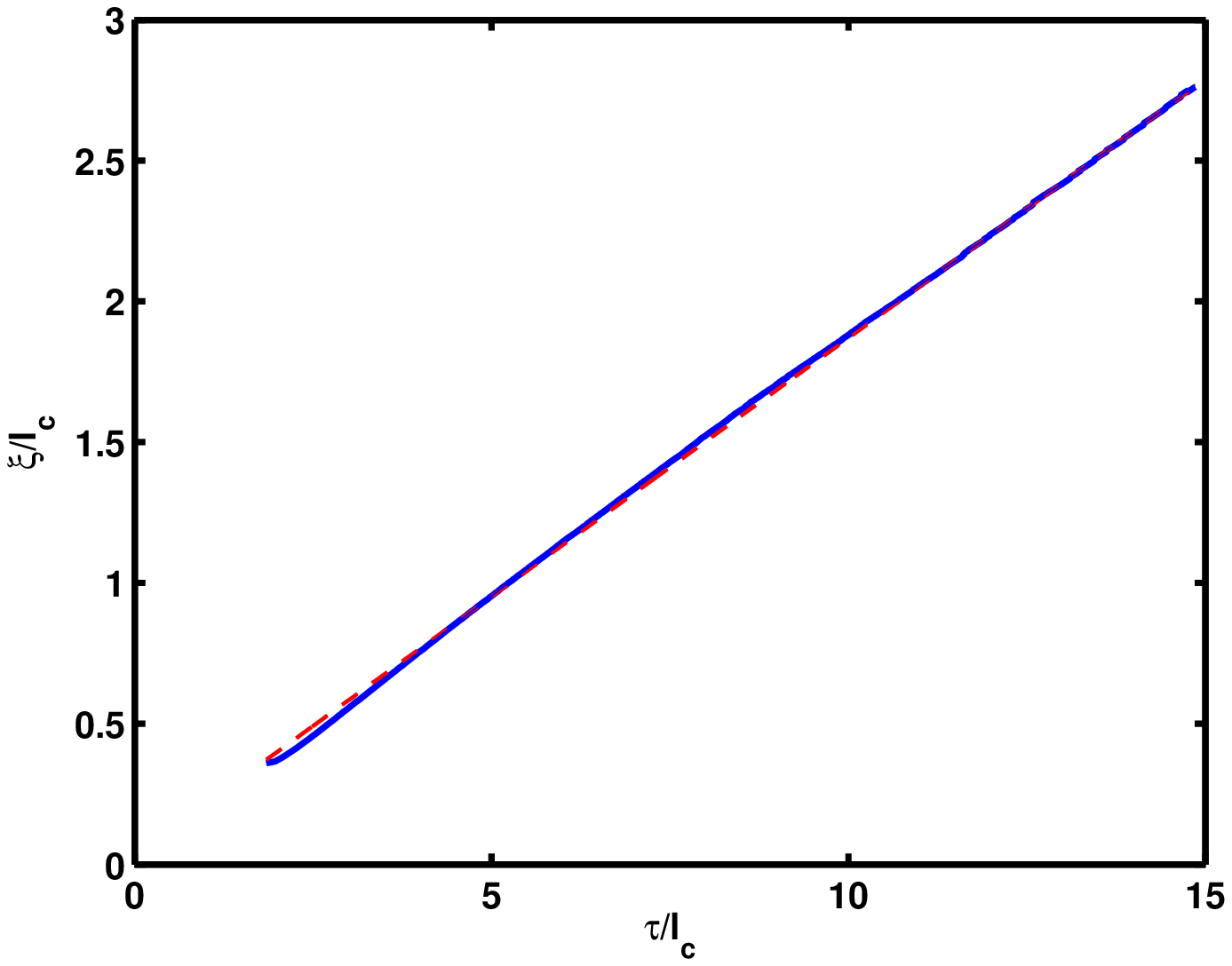}

\caption{\label{f:NGRin}  Left: snapshot from a Nambu-Goto string simulation, showing loops formed during the relaxation from the initial state (blue) and during scaling (red). Right:
infinite string comoving length scale $\xi$ (as computed from the invariant length $L = \int \ep d\si$) for a Nambu-Goto simulation 
\cite{Ringeval:2005kr}.
Distance and conformal time are given in terms of the initial comoving correlation length $l_c$, which is the only fixed length scale in the system (snapshot and data courtesy C.\ Ringeval).}

\end{figure}

When strings curl back on themselves and reconnect, they produce a loop, and creating a pair of oppositely-moving kinks (discontinuities in the string tangent vector $\bX'$) both on the loop and the parent string.  Kinks are an important feature of string networks: classical Nambu-Goto network simulations 
\cite{OrigNGsim,Ringeval:2005kr,OluVan}
show that strings are full of kinks (see Fig.\ \ref{f:NGRin}) left behind by loops. This loop production allows the infinite string length density to scale, although the total length of string is fixed by conservation of energy.

Note that the slope of $\xi$ is lower than that for Abelian Higgs strings, meaning that the infinite string length density is higher, by about a factor of 2.5  
\cite{Hindmarsh:2008dw}. However, the main difference is clearly the presence of string loops in the Nambu-Goto simulations.  The visual impression in Fig.\ \ref{f:NGRin} is dominated by the loops produced earliest, during the relaxation to scaling, which disguises fact that the loops are produced in such a way as to maintain a scaling power law distribution for loop size.  

Scaling arguments show that the number of (stable, non-self-intersecting) loops with physical length in the interval $[l,l+dl]$ is 
\cite{Vachaspati:1984dz}
\ben
n(l,t) = \horDis^{-4}n(l/t), \qquad n(x) \propto x^{3\nu - 4}, \quad l \gtrsim l_*=a(t)l_c
\een
where the scale factor grows as $a \propto t^\nu$,  $l_c$ is the initial comoving correlation length, and $\horDis$ is the horizon distance.\footnote{Departures from the scaling power law indices were observed in \cite{Ringeval:2005kr}, 
which was suggested in \cite{BlancoPillado:2011dq} to arise from loop production at small scales, or to be a sign that the loop distribution was not yet scaling.}


One can define a loop production function by 
$
f(l,t) = a^{-3} \frac{\pa}{\pa t}\left(a^3n(l,t)\right),
$
and express it in scaling form 
\ben
f(l,t) = \horDis^{-5} f(x), \quad f(x) \propto x^{-p}. 
\een
Recent simulations \cite{BlancoPillado:2011dq} are uncovering this scaling form, which takes a surprisingly long time to emerge, and show a very broad distribution consistent with $p\simeq 2$ over a couple of orders of magnitude in $x$, with broad peaks at $x \simeq 0.05$ and at the lower cutoff $l_*=a(t)l_c$ set by the initial comoving correlation length $l_c$.  One can show that $\int x f(x) dx$ is bounded by conservation of energy, and hence $p < 2$.  
%
The true loop distribution must take into account the decay of loops into radiation, which modifies the simple power laws \cite{LoopDistModels}. 

The difference in the loop number densities between Nambu-Goto and
Abelian Higgs simulations demands some explanation.  If one is treating
the Nambu-Goto action as an approximation to the motion of solitonic
strings, one is operating beyond its strict domain of validity (large
curvature radius) as the string is full of kinks where the curvature
radius $R$ is of order the string width $w$. There is therefore no
particular reason to expect the behaviour to be the same. In the
Nambu-Goto approximation, the massive modes of the field are eliminated
from the outset, and so infinite Nambu-Goto strings can lose energy only
by producing loops. However, it is puzzling that Abelian Higgs strings
with $w/R \ll 1$ can produce radiation with frequency $\om \sim w^{-1}
\simeq e\phi_0$, while naively it appears to be 
made of fields changing with a frequency $R^{-1} \ll w^{-1}$.  
Indeed, for a smooth string prepared in a standing wave of wavelength $\sim R$ the power per unit length in massive radiation goes as  $dP/dL \sim (\mu/R)\exp(-R/w )$ \cite{Olum:1999sg}, and massive radiation becomes negligible as $R\to\infty$. 
Thus it is often argued that a network of Abelian Higgs strings must eventually start behaving like Nambu-Goto strings.  
However, real string networks do not prepare their initial conditions so carefully, and massive radiation continues at a rate per unit length $dP/dL \sim (\mu/R)$ in the largest simulations to date, where $w/R \sim 10^{-2}$.  In 2D domain wall networks massive radiation operates at this rate at least until $w/R \sim 10^{-3}$ \cite{Borsanyi:2007wm}, so there is clearly a mechanism at work which cannot be ignored. While the mechanism is not well understood, small scale structure on infinite strings (discussed in the next section) has been shown to be important\cite{Hindmarsh:2008dw}, and such energy cascades from large to small scales are familiar from the study of turbulence.

Thus it is an open question whether Nambu-Goto strings approximate the motion of solitonic strings at late times, and in particular whether soliton strings produce any significant population of loops.  This has an important bearing on all observational signals from loops.

For macroscopic elementary strings the situation is also  unclear as the technology to do a proper quantum calculation is not well evolved: it is difficult enough even to construct the states \cite{BlancoPillado:2007hi,Skliros:2009cs}. However,  
given that the low-energy approximation for a macroscopic elementary
string is just the classical Nambu-Goto action, an intriguing
possibility is that Nambu-Goto simulations are actually modelling cosmic
superstrings rather than their solitonic counterpart.  If this is true,
it should be quite easy to tell solitonic and elementary strings apart
based on observations of 
the loop population alone.

\subsection{Small scale structure}
\label{ss:SSS}

Early Nambu-Goto string simulations  \cite{OrigNGsim} showed that there were ``wiggles" on infinite strings, excitations with wavelengths much less than the network
length scale $\xi$. The wiggles took the form of closely
spaced kinks, with as much as 50\% of the energy of the infinite
string contained in these high frequency waves. Characterising the wiggles by their fractal dimension $d_f$ on a scale $l$, numerical simulations show that $d_f$ rises gradually from 1 on scales small compared with $\xi$ (approximately straight strings) to 2 on large scales (Brownian random walks) \cite{Martins:2005es}.

This small scale structure can also be characterised by the tangent vector correlation function, which goes as 
\cite{Polchinski:2006ee}
\ben
C_T(\si,t) = \vev{\bXp(\si)\cdot\bXp(0)} = \vev{\bXp(0)^2}\left(1 - A(\si/t)^{2\chi} \right),
\een
where $A$ is an O(1) constant and the exponent $\chi$ determined from the mean square velocity $\bar v^2$ and the Hubble expansion exponent $\nu$ as 
\ben
\chi = \frac{\nu(1-2\bar v^2)}{1 - \nu(1-2\bar v^2)}.
\een
This scaling power law form for small scale structure has been found both on Nambu-Goto \cite{Polchinski:2006ee} and Abelian Higgs strings \cite{Hindmarsh:2008dw}.  At the very smallest scales, the correlation function approaches a constant as $\si$,  explicable in terms of the kinks \cite{Copeland:2009dk}.  The kink separation decreases very slowly, and scaling is absent.


The significance of the small scale structure is that it is involved in transporting energy from large to small scales on infinite strings.  
The small scale structure triggers loop production near cusps (points on the string where the tangent vector vanishes and the string moves at the speed of light) \cite{Polchinski:2006ee}. On Abelian Higgs strings it contains the necessary high frequencies for producing massive radiation \cite{Hindmarsh:2008dw}.

\subsection{Gravitational radiation reaction}
\label{ss:RadReact}

Early string simulations \cite{OrigNGsim} showed no sign of the expected scaling in the loop distribution.  The belief in scaling was strong, and it was conjectured that the typical loop size was set by gravitational radiation damping of small amplitude waves on infinite strings \cite{OrigNGsim}.  In the radiation reaction scenario the relevant scale for loop production is $\vev{l} \sim (G\mu)^pt$, with $p$ variously estimated to be 1, \cite{Hindmarsh:1990xi}
$(1+2\be)/2$ \cite{Siemens:2002dj}, and $1+2\chi$ \cite{Polchinski:2006ee}, where $\be = 1(2)$ in the matter (radiation) dominated era and $\chi$ is the exponent of the power spectrum of small-scale structure on infinite strings.

A numerical check of gravitational backreaction \cite{Quashnock:1990wv} demonstrated the expected damping of high frequencies on a simple kinky loop. However, no network simulation includes 
gravitational radiation reaction.  In particular, it is not clear how 
damping operates on a scaling network which is self-intersecting and producing kinks.  

\subsection{Summary}

There are two scenarios for ordinary cosmic strings without junctions, which we refer to as the Nambu-Goto and Abelian Higgs scenarios, to make clear their origin.

Both recognise that the length density (both comoving and physical) in infinite strings scales, that is $L\tau^2/V$ is a constant. In the Abelian Higgs scenario this is about 2.5 -- 3 times as high as in the Nambu-Goto scenario (see Table, after \citen{Bevis:2010gj}). 

\begin{table}
\centering

\begin{tabular}{|l|l|c|c|}
\hline
Simulation & Measure                           & Radiation     & Matter \\
\hline
AH \cite{Bevis:2010gj} & $\xi/\tau$                 & $0.255\pm0.018$ & $0.285 \pm 0.011$ \\
AH \cite{Bevis:2010gj}& $L\tau^2/V$                & $15\pm2$        & $12.2 \pm 0.96$ \\ 
\hline
NG \cite{Ringeval:2005kr}  & $L\tau^2/V$  & $48$ & $27$ \\
NG \cite{Martins:2005es} & $L\tau^2/V$  & $37.8 \pm 1.7$ & $28.4\pm 0.9$ \\
 NG \cite{BlancoPillado:2011dq} & $L\tau^2/V$  & $44$ & $35$ \\
\hline

\end{tabular}

\caption{\label{t:StrDen} Numerical results for the network length scale in
horizon units and the string length density $L/V$ normalised
to the horizon size $\tau$. AH labels results from
Abelian Higgs simulations, while NG labels 
results from Nambu-Goto simulations. Note the NG
results quoted includes only infinite strings, and that not all authors include an error estimate.  See \cite{Hindmarsh:2008dw} for a discussion of the AH string length measure.}

\end{table}

In the AH model, strings lose their energy primarily into scalar and gauge radiation.  There is O(1) horizon-sized loop produced per Hubble time, and the loops have a lifetime of order the Hubble time, decaying into radiation just as the infinite strings do.

In the NG model, strings lose their energy into loops, whose sizes range between the comoving initial correlation length and about a tenth of the horizon size $\horDis$. The average loop size at formation is the comoving initial correlation length. On scales between the  comoving initial correlation length and $0.1\horDis$ the rate at which  energy (length) is emitted into loops is approximately constant per unit logarithmic length interval.

Gravitational radiation reaction is widely believed to play a critical role in setting the lower cut-off on a real string network.  In view of the importance of establishing the typical loop size at formation for many observables, particularly gravitational radiation background from strings, a simulation including radiation reaction is required.

Finally, it should be noted that we are extrapolating from the end of inflation ($t_i\sim 10^{-36}$ s) to today ($t_0 \sim 3\times 10^{17}$ s). Over that range of scales even logarithmic corrections become O(100), so acquiring a better theoretical understanding of scaling in string networks, perhaps by techniques taken from other fields \cite{Hindmarsh:1996xv},  is extremely important.  

\section{Observational signals of cosmic strings}

\subsection{Stochastic background of gravitational radiation}
\label{ss:GravRad}

An oscillating loop of string is the source of gravitational radiation with total power \cite{Vilenkin:1981bx,Burden:1985md}
\ben
P = \Ga G\mu^2,
\een
where $\Ga \sim 10^2$ is a dimensionless constant depending on the loop trajectory. The radiation power of a loop of length $l$ is peaked at the fundamental frequency $\om_1 = 4\pi/l$, and for high harmonics $\om_n$ decreases as $n^{-4/3}$, due to the strongly beamed radiation from cusps.  Both the stochastic background from the integrated sum of all loops on our backward lightcone \cite{Vilenkin:1981bx,Hogan:1984is,Caldwell:1991jj}, and the rare but strong signals from individual cusps \cite{Damour:2000wa} offer prospects for detection. 


Let us denote the average size at formation by $\ell_{\rm i} = \al t$, and also to scale out a factor $G\mu$ and define $\ep = \al/G\mu$. In the original radiation reaction scenario $\ep$ was a constant independent of $G\mu$, but in more recent versions it is a higher power of $G\mu$.

There are limits on the stochastic background from millisecond pulsar timing \cite{Hogan:1984is}, from Big Bang Nucleosynthesis \cite{Davis:1985pu}, and from LIGO \cite{Abbott:2009ws}.  The strongest limits are from the pulsar timing, giving \cite{Jenet:2006sv}
\ben
\Om_gh^2 < 2 \times 10^{-8}
\een
In the Nambu-Goto model, the stochastic background from string loops is, for $\al \ll \Ga G\mu$, \cite{Caldwell:1991jj}
\ben
\Om_gh^2 \sim 10^{-4}\frac{G\mu}{x_*^2\Om_{\rm m}},
\een
where $x_* = \xi_{\rm p}/\horDis$ and $\Om_{\rm m}$ is the matter density parameter today.  Hence one can estimate $G\mu \lesssim 10^{-6}$.  A more accurate analysis \cite{Battye:2010xz} gives
\ben
G\mu < 7 \times 10^{-7}
\een
As there is only O(1) loop per Hubble volume in the Abelian Higgs model, there are no constraints from gravitational radiation.



\subsection{Cosmic rays from cosmic strings}
\label{ss:CosmicRays}

Solitonic strings can decay into the particles of the fields from which they are made. There are three main avenues: perturbative production \cite{Vachaspati:1984yi,Srednicki:1986xg}; cusp annihilation \cite{Brandenberger:1986vj,BlancoPillado:1998bv}; and non-perturbative massive radiation \cite{Vincent:1997cx,Olum:1999sg}.  

For loops found in network simulations of size $l$ and width $w \sim m_{\rm v}^{-1}$, the power in the three mechanisms scales as
\begin{enumerate}

\item Perturbative particle production: $P \sim \mu (w/l)$

\item Cusp annihilation: $P \sim \mu (w/l)^{1/2}$

\item Massive radiation: $P \sim \mu $.

\end{enumerate}

The largest flux, and therefore the most important bounds, comes from non-perturbative massive radiation, assuming that the scaling configuration observed in AH simulations continues indefinitely.  Cosmic strings produce massive particles, denoted $X$,  at a rate 
\ben
 \dot n_X \simeq \frac{Q_0}{m_X}\left( \frac{t}{t_0}\right)^{-3},
\een
where $t_0$ is the time today, and the energy injection rate $Q_0$ is given by 
\ben
Q_0 = - \dot\rho_{\rm s}(t_0) = \frac{6\mu}{x_*^2\horDis(t_0)^3}. 
\een
Cosmic strings constitute a $p=1$ topological defect (TD)\cite{Bhattacharjee:1998qc}  model, for which  
there is an 
upper bound on the energy injection rate from the low energy diffuse $\ga$-ray background \cite{Bhattacharjee:1998qc} 
\ben
Q_0 \lesssim 4.4\times 10^{-23}h \; \textrm{eV cm$^{-3}$ s$^{-1}$}.
\een
Assuming the massive particles decay into Standard Model particles with a non-zero branching fraction $f$, one can derive a bound \cite{Bhattacharjee:1997in,Wichoski:1998kh}\footnote{Note that these authors give a bound a factor of 10 lower, as they took a value of the string density approximately 10 times higher than indicated in current Abelian Higgs simulations.}
\ben
G\mu \lesssim 10^{-10} f^{-1}.
\een
Although the strings produce particles with energies up to $m_X$, which may be up to $10^{16}$ GeV, there is no bound from Ultra-High Energy Cosmic Rays (UHECRs, defined as cosmic rays with energies greater than $10^{20}$ eV). The range of these particles is less than 100 Mpc, and the nearest string in the solitonic string scenario of order a Hubble distance\footnote{The bound derived in Ref.\ \citen{Vincent:1997cx} from the UHECR flux is therefore incorrect.}


If strings are described by the Nambu-Goto model, 
there is no significant bound from cosmic rays. Hence cosmic ray and gravitational radiation bounds are complementary.

\subsection{Cosmic Microwave Background constraints}
\label{ss:CMBconstraints}


The gravitational field of a moving string between us and the last scattering surface produces a discontinuity in the apparent  temperature of the Cosmic Microwave Background (CMB), \cite{GKS}
\ben
\De T \simeq 8\pi (G\mu)vT_\mathrm{CMB},
\een
where $v$ is the transverse speed (see Figure.\ \ref{f:CMBmapsSmallScale}).  Hence $(\De T/T) \simeq 10^{-5} (G\mu/10^{-6})$, which means that GUT or inflation-scale strings produce observationally interesting CMB perturbations. 
To resolve the distinctive edges of the 
Gott-Kaiser-Stebbins (GKS) 
effect one needs both high resolution and high sensitivity.  A number of limits can be derived by directly looking for these edges: 
from OVRO \cite{Hindmarsh:1993pu}
\(G\mu < 11\times10^{-6} x_* 
\), where $x_* \simeq 0.3$ is the string length scale in units of the horizon, 
and from WMAP \cite{Jeong:2004ut}
\(G\mu < 3.7 \times 10^{-6}\).
See also Refs.\ \citen{Lo:2005xt,Amsel:2007ki} for other edge detection methods.

More stringent constraints come from calculations of the CMB temperature and polarisation power spectra and their comparison to the data, principally WMAP \cite{Wyman:2005tu,Battye:2006pk,Bevis:2006mj,Battye:2010xz}.  To calculate power spectra, one solves the Boltzmann equation for the photon-baryon fluid  coupled to the linearised Einstein equations in an expanding background, with an additional source in the form of the energy-momentum of the cosmic strings  (see e.g.\ Ref.\ \citen{Hu:1997mn}). Schematically, the Fourier-space perturbation equations can be written  
\ben
\D_{\al\be}(\tau,k) h_\be(\tau,k) = S_\al(\tau,k),
\een
where \( h_\al(\tau,k)\) are the perturbations in the metric, matter, radiation intensity, or polarisation;  
\( D_{\al\be}(\tau,k)\) is a (time dependent) time evolution operator; and  
\( S_\al(\tau,k)\) is the source. These equations can be re-expressed as coupled evolution equations for the multipole moments of the CMB angular power spectrum. In the standard cosmological model,  inflation determines the initial conditions for $h_\be(\tau,k)$ and there are no sources. Standard Einstein-Boltzmann solvers \cite{EBsolvers} evolve the initial conditions forward and deliver the CMB angular power spectra, and also the matter power spectrum.


The CMB angular power spectra are derived from the multipole moments of the temperature and polarisation distributions.  For example, for the temperature in direction $\bn$ one defines 
\(a_{lm} = \int d\Omega \De T(\bn) Y^*_{lm}(\bn) \), and the angular power spectrum is 
\ben
C_l = \sum_{m=-l}^l |a_{lm}|^2.
\een
Most often plotted is the anisotropy power, 
\( l(l+1)C_l/(2\pi) \).
On a linear-log plot the power over a given multipole range is proportional the area under the curve.


The standard techniques can be adapted to include sources.
The contribution to the power spectra from the sources can be written schematically as \cite{Turok:1997gj,Durrer:1998rw}
\ben
\vev{|h_\al(\tau_0,k)|^2} = \int \int \D^{-1} \D^{-1}  \vev{S_\al(\tau,k)S^*_\al(\tau',k)}
\een
where the angle brackets indicate an average over an ensemble of source histories, and we denote the numerical integration symbolically by $ \D^{-1}$. There are two approaches to computing the averages.

\begin{figure}

\centering

\includegraphics[width=0.7\textwidth]{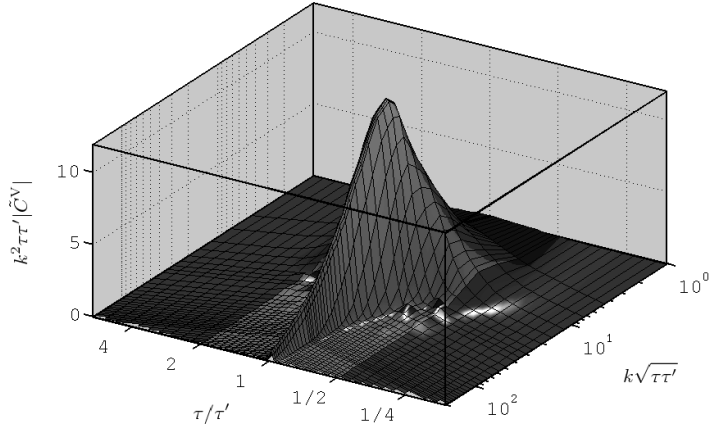}

\caption{\label{f:UETCvec} The vector unequal time correlator (UETC) from a cosmic string simulation \cite{Bevis:2010gj}.  Light grey comes from an average over data from three simulations, dark grey indicates extrapolation.}

\end{figure}

In the unequal-time correlator (UETC) method one performs a number of numerical simulations, and averages to find the UETCs of the source energy momentum tensor
\[
C_{\mu\nu\rho\lambda}(k,\ta,\ta') = 
\left\langle T_{\mu\nu}(k,\ta) T_{\rho\lambda}^{*}(k,\ta') \right\rangle.
\]
Due to the rotational symmetry of the system, there are in fact only 5 independent UETCS: 3 scalar, 1 vector, and 1 tensor. 
Furthermore, scaling makes the correlation functions behave as 
\ben
C(k,\ta,\ta') \sim \phi_0^4\frac{z^n}{\sqrt{\ta\ta'}} \tilde C(z,r),
\een
where $r = \ta/\ta'$, $z = k \sqrt{\ta\ta'}$, and $n=2$ is conventionally chosen for the vector correlator.  The dimensionless correlators $\tilde C$  are peaked at $r=1$, $z\sim1$ with a ridge at $r=1$ of width O(1) for $z \lesssim 1$, and $1/z$ for $z \gtrsim 1$.  
In practice scaling is vital to making the calculation feasible: 
simulations over only a few expansion times are required.  Figure \ref{f:UETCvec} shows the vector UETC from an Abelian Higgs cosmic string simulation \cite{Bevis:2010gj}.

The UETCs are symmetric under interchange of $z$ and $z'$, and therefore can be decomposed into orthogonal 
eigenvectors $\psi_\la(k,\tau)$ of the UETCs, defined from 
\ben
\int_{x_{\rm i}}^{x_0} dx' \tilde C(x,x') \psi_\la(x') = \la \psi_\la(x),
\een
where $x = k\tau$, $x_0$ is the value of $x$ today, and $x_{\rm i} \ll 1$. One can then apply the Einstein-Boltzmann solver to the eigenvectors, and recover the power spectra by summing over the separate contributions. 

\begin{figure}
\centering

\includegraphics[width=0.54\textwidth]{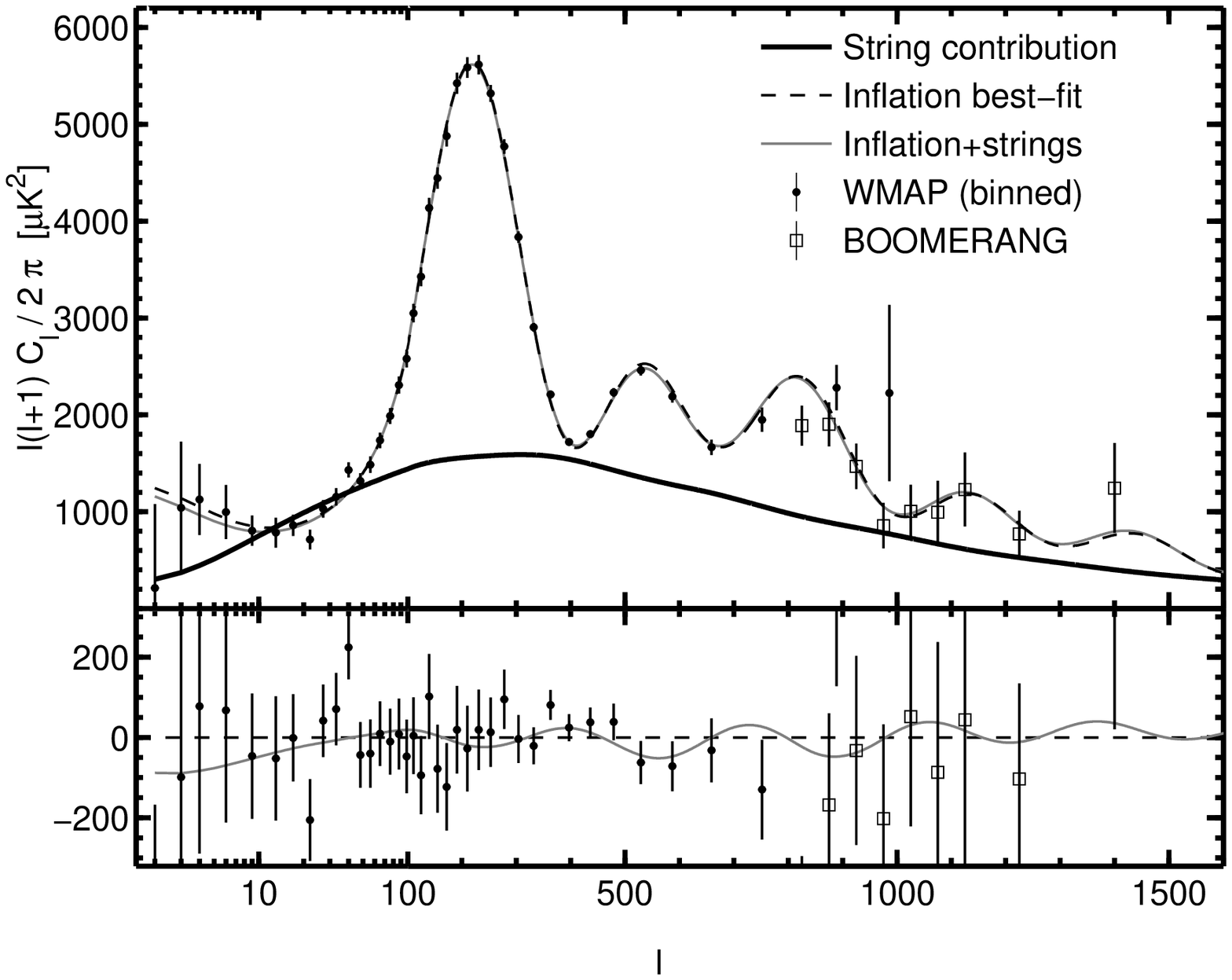}
\includegraphics[width=0.44\textwidth]{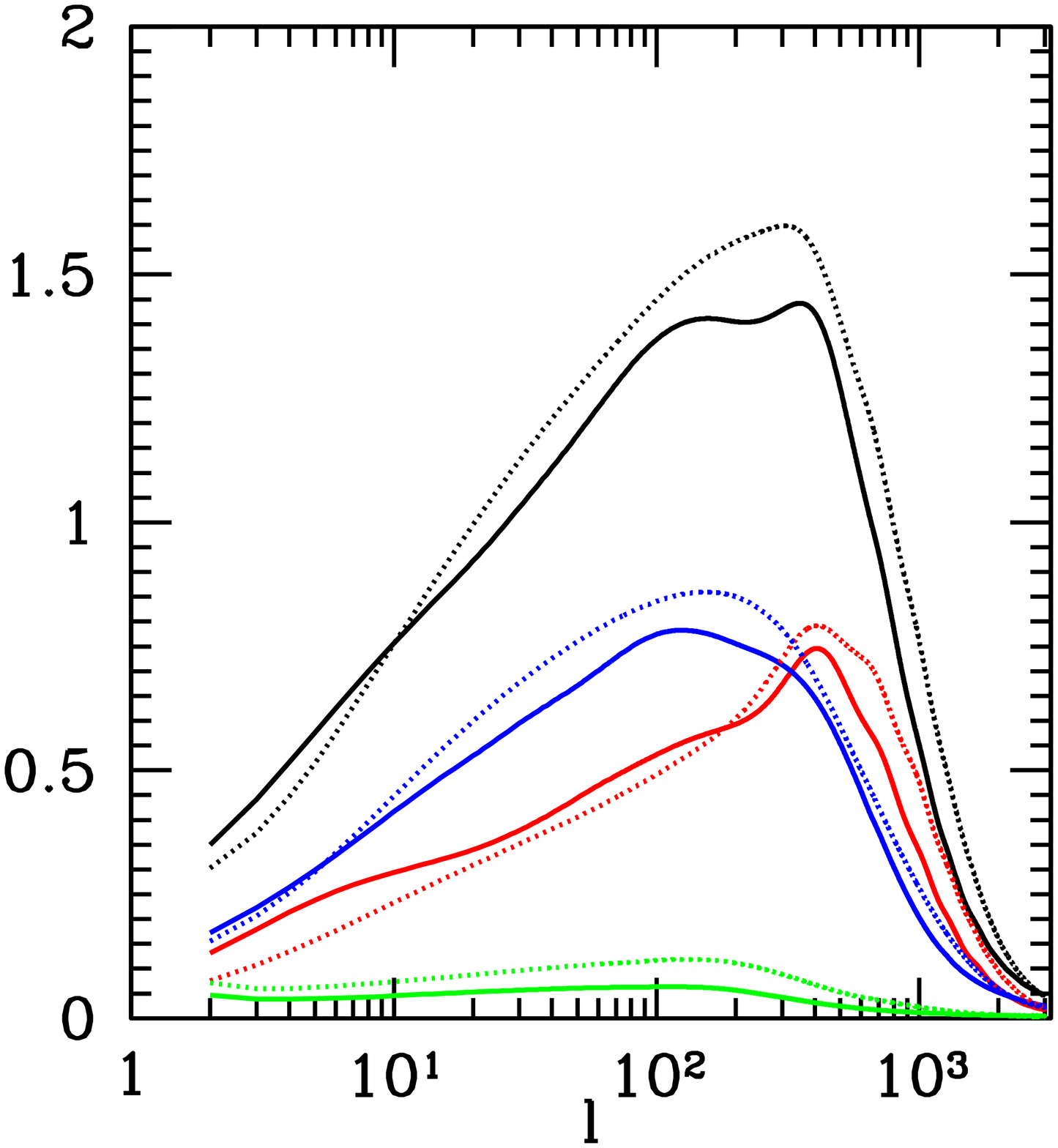}

\caption{\label{f:TTstrInf} {Left top:} Strings normalised to WMAP3 ($\ell=10$) \cite{Bevis:2006mj}
{Left bottom:} Differences from best-fit standard $\La$CDM  cosmology. Right:  
\label{f:USMcompare} Comparison of scalar (red), vector (blue) and tensor (green) power spectra in the Unconnected Segment Model (solid) with those produced from the UETCs of the Abelian Higgs model simulations (dotted) \cite{Battye:2010xz}.}

\end{figure}

Another approach is to model the sources more economically than a full 3D simulation of the field equations, in such a way that the UETCs are reproduced.  For strings, the Unconnected Segment Model  \cite{Vincent:1996rb,USM}
models the strings as moving sticks of energy  $\mu$, tension $T$ and length $L(t) = x_* t$.  The sticks have random positions and velocities $\bar v$, and decay at random times so that the energy density decays as $t^{-2}$.  It has four parameters: $\mu$, the ratio $\be = T/\mu$, $x_*$ and $v$, which can be related by a phenomenogical model such as the velocity-dependent one-scale (VOS) model \cite{Martins:2000cs}.  There is enough freedom to model either the Abelian Higgs or Nambu-Goto simulations \cite{Battye:2010xz} (see Fig.\ \ref{f:USMcompare}).  Nambu-Goto simulations are modelled with larger $\be$ and $v$, and smaller $x_*$, and tend to have a relatively smaller vector contribution.

Armed with the power spectra, one can apply standard Markov Chain Monte Carlo (MCMC) techniques (see e.g.\ Ref.\ \citen{Lewis:2002ah}) to fit models with both inflation and cosmic strings to the data \cite{Bevis:2006mj}.  The two sources of perturbations are not correlated: they are produced by fluctuations in different fields at different times, and so the power spectra can be simply added. 
The cosmic string power spectra are proportional to $(G\mu)^2$, so this results in its simplest form in a cosmological model with one more parameter, adding to the standard set $\Om_b h^2$ (fractional baryon density), $\Om_c h^2$ (fractional Cold Dark Matter density), $\theta_A$ (acoustic scale), $\powspec(k_0)$ (scalar amplitude at the pivot scale $k_0$), and $n_s(k_0)$ (scalar tilt). One can also usefully parametrise the string contribution by its fractional contribution to the temperature angular power spectrum at a multipole $\ell = 10$, \(f_{10} = C_{10}^\mathrm{TT,string}/C_{10}^\mathrm{TT,total}\).


One finds a degeneracy between $G\mu$ and $n_s$ (or $\Om_bh^2$), and it was found that the best fit to the CMB data (3-year WMAP, with further CMB data on small scales) was a mixed inflation and AH string model with $f_{10} = 0.099$ with $n_s = 1.00$ \cite{Bevis:2006mj}.  The degeneracy can be lifted by fitting to data such as the SDSS luminous red galaxy (LRG) power spectrum \cite{Tegmark:2006az}, or applying priors from Hubble parameter measurements \cite{Riess:2009pu}, Baryon Acoustic Oscillations \cite{Percival:2009xn}, and deuterium abundance measurements \cite{Pettini:2008mq}.  The CMB remains the most robust from both the observational and theoretical point of view. 

The most recent results from fits to the CMB alone give $G\mu < 6.8 \times 10^{-7}$ for the Abelian Higgs model and  $G\mu < 2.8 \times 10^{-7}$ for the Nambu-Goto model \cite{Battye:2010xz}, using the 5-year WMAP data release in combination with CMB data on smaller scales.  Combining the 7-year WMAP data with recent ACT data gives  $G\mu < 1.6 \times 10^{-7}$ for the NG model \cite{Dunkley:2010ge}.

%
%
%
%
%
%
%
%
%
%
%
%

\begin{figure}

\centering

\includegraphics[width=0.5\textwidth]{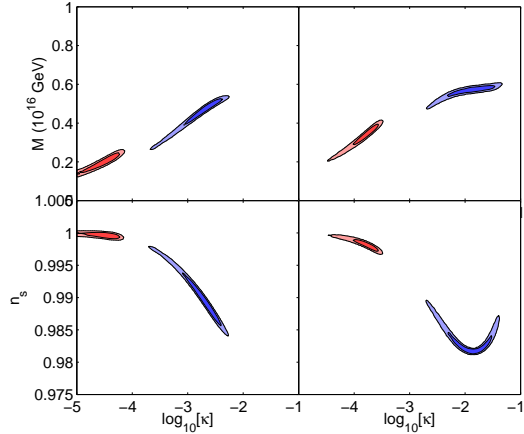}

\caption{ \label{f:FDfit} 
Marginalised likelihoods for D-Term (red, smaller values of $\ka$)
and F-term (blue, larger values of $\ka$) models of Section \ref{ss:InfStrings}. The string models are for Nambu-Goto (left pair) and Abelian Higgs (right pair) using CMB and SDSS data (from the second of Refs.\ \citen{Battye:2006pk}).
%
}
\end{figure}

In hybrid inflation models the string tension and the scalar amplitude
are related by the parameters in the potential, which can therefore be
directly constrained. In the minimal D-term and F-term inflation models
the important parameters are the scalar coupling $\ka$ and the
symmetry-breaking scale $M$ (see Eq.\ (\ref{e:superpot})
).  Figure \ \ref{f:FDfit} 
shows the $1\si$ and $2\si$ likelihood contours derived from a MCMC fit to CMB data (including 5-year WMAP) and SDSS LRGs \cite{Battye:2006pk}.  
The best fit models give $\Obhh$ in the range $0.024$ -- $0.025$, which is in tension with the value inferred from the deuterium abundance $D/H$ in damped Lyman-$\al$ systems, $\Obhh = 0.0213 \pm 0.0010$ via light element abundances predicted by standard Big Bang Nucleosynthesis (BBN) \cite{Pettini:2008mq}.  However, given that there are only 7 measurements, with abundances corresponding to $\Obhh$ in the range  $0.017$ to $0.030$, the BBN constraint should be treated with caution.

Future CMB data will be sensitive to lower values of $G\mu$. Planck is scheduled report its power spectra in early 2013, and there are plans for a dedicated satellite missions to measure accurately the B-mode of the polarisation power spectra.
\cite{Baumann:2008aq,Collaboration:2011ck}

An important question is whether a string contribution can be distinguished from other departures from the standard adiabatic scenario, such as gravitational waves (predicted in large field inflation models), and other kinds of topological defects such as textures and semilocal strings \cite{Urrestilla:2008jv,Mukherjee:2010ve}.
 
\begin{figure}

\centering

\includegraphics[width=0.5\textwidth]{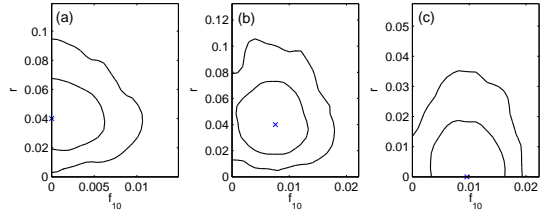}

\caption{\label{f:PlaF10R} Likelihood contours (68\% and 95\%) of the marginalized 2D posterior distribution for a cosmological model with (a) tensor-to-scalar ratio $r = 0.04$ and no string component, (b) r = 0.04 and defect fraction $\fd = 0.008$, and (c)$\fd = 0.01$ with no tensor component.\cite{Urrestilla:2008jv} The actual models are depicted with a cross.
}
\end{figure}

It is conventional to parametrise the gravitational wave contribution in term of the tensor-to-scalar ratio \(r = \powspec_t/\powspec_t\). 
From Fig.\ (\ref{f:PlaF10R}) one can infer that Planck can distinguish a string model \(\fd \gtrsim 0.02\) from inflationary tensor fluctuations \cite{Urrestilla:2008jv}.  The reach in $\fd$ for a future B-mode satellite depends on assumptions made about foregrounds and the efficiency of separating out the contribution from the lensed E-mode power spectrum. With conservative assumptions, the B-mode missions will be sensitive to \(\fd \gtrsim 2\times10^{-3}\) \cite{Mukherjee:2010ve}.  Under the assumption of perfect delensing, one can reach 
\(\fd \gtrsim 10^{-5}\) \cite{Seljak:2006hi,GarciaBellido:2010if}.

%
%
%
%
%



\subsection{Cosmic string CMB non-Gaussianity}

The GKS effect introduces a highly non-Gaussian signal at small angular scales, where the string signal is expected to dominate the power spectrum.  

Let $\Theta = \Delta T/T$.  In the flat sky approximation, and in the light cone gauge $X^0 + X^3 = \tau$, one can show \cite{Stebbins:1987va,Hindmarsh:1993pu}
\[
-k^2\Theta_\bk = i8\pi G\mu k^A   \int d\si \Xd^A(\si)
e^{\textstyle i\bk\cdot\bX(\si)},
\]
where $A=1,2$ are the 2D transverse coordinates.

\begin{figure}

\centering

\includegraphics[width=0.45\textwidth]{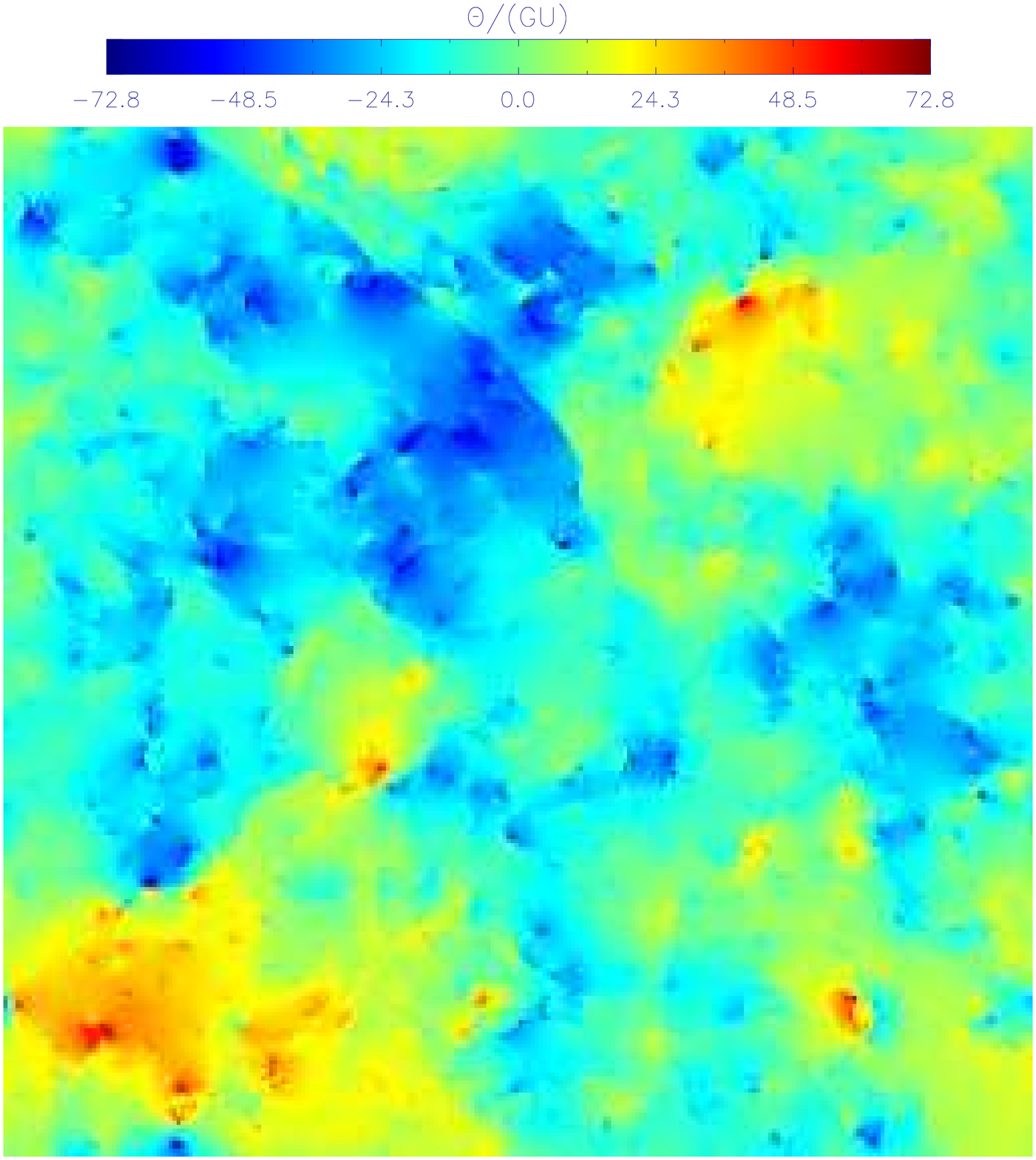}
\hfill
\includegraphics[width=0.45\textwidth,angle=90]{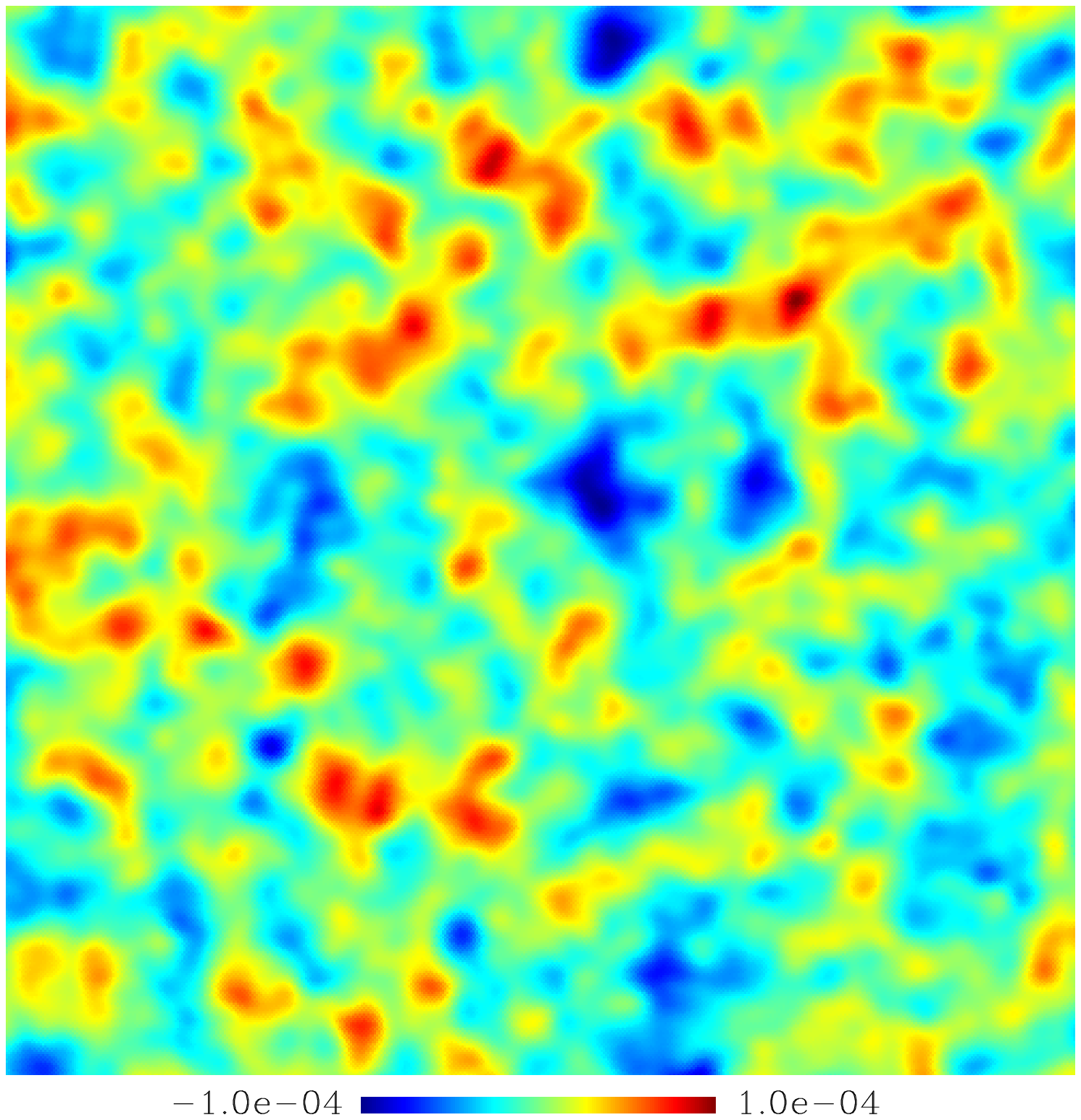}

\caption{\label{f:CMBmapsSmallScale}  Small angular scale maps of the CMB fluctuations induced by cosmic strings.  Left: GKS effect only, size $7.2^\circ$, resolution $0.42''$ \cite{Fraisse:2007nu}. Right:  acoustic oscillations included, size $3^\circ$, resolution resolution $0.70'$  \cite{Landriau:2010cb}.}

\end{figure}

By taking an average over an ensemble of strings, either analytically or numerically, one can show that 
the resulting power spectrum has an approximately $\ell^{-1}$ tail \cite{Bouchet:1988hh,Hindmarsh:1993pu,Fraisse:2007nu,Yamauchi:2010ms}, which dominates the string-induced acoustic oscillations for \(\ell\gtrsim 3000\) or angular scales of less than about $3'$ \cite{Pogosian:2008am,Bevis:2010gj}.  However, calculations which include the acoustic oscillations 
with an Einstein-Boltzmann solver
\cite{Landriau:2010cb} show that they 
may
tend to obscure the GKS edges (see Fig.\ \ref{f:CMBmapsSmallScale}).

One can also work out the CMB bi- and trispectrum at small angular scales from the GKS effect by averaging over string ensembles \cite{Hindmarsh:2009es,Regan:2009hv}. The bispectrum $b_{\bk_1,\bk_2,\bk_3}$ and the trispectrum $T_{\bk_1,\bk_2,\bk_3,\bk_4}$ are defined by
\bea
 \vev{\Theta_{\bk_1}\Theta_{\bk_2}\Theta_{\bk_3}} &=&
b_{\bk_1,\bk_2,\bk_3}(2\pi)^2\de(\bk_1+\bk_2+\bk_3), \\
 \vev{\Theta_{\bk_1}\Theta_{\bk_2}\Theta_{\bk_3}\Theta_{\bk_4}} &=&
T_{\bk_1,\bk_2,\bk_3,\bk_4}(2\pi)^2\de(\bk_1+\bk_2+\bk_3+\bk_4),
\eea
and one finds that the GKS effect gives an equilaterial bispectrum of order 
\begin{equation}
b_{kkk} \simeq  -(4\times 10^{-14})c_0
 \frac{\vb^2}{\tb^4}\frac{L\hat\xi}{\cal A}
 \frac{1}{\hat\xi^2k^6} \left(\frac{G\mu}{0.7\times 10^{-6}}\right)^3.
\end{equation}
where $\bar v$, $\bar t$, $L\hat\xi/{\cal A} $, and $c_0$ are dimensionless constants characterising the string network.  
All but $c_0$ are O(1), with $c_0 \sim 10^{-1}$.
\footnote{The smallness of $c_0$ can be understood as the result of approximate time reversal invariance for string motion on small scales (B. Wandelt, private communication, see also Ref.\ \citen{Hindmarsh:2009qk}).}
The GKS bispectrum (and therefore the skewness) is negative, and of order $C_\ell^{3/2}$.
It is not straightforward or unambiguous translating the predictions for the bispectrum into an equivalent local non-linearity parameter $f_{nl}$ \cite{Komatsu:2001rj}. Using the relation \(b_{lll} \simeq (2\times 10^{-17})l^{-4}  f_{nl}\) \cite{Komatsu:2001rj}, one finds that $|f_{nl}| \sim 10^3c_0(G\mu/0.7\times10^{-6})^3$ at $k\hat\xi\sim1$ or $\ell \simeq 1000$. A comparison using a sum over multipoles gives $|f_{nl}| \sim 75c_0(G\mu/0.7\times10^{-6})^3$ \cite{Regan:2009hv}.

The trispectrum goes as 
\ben
T_{\ell\ell\ell\ell} \sim (G\mu)^4\ell^{-\rho}, 
\een
with $\rho = 6(1 + 2\chi/3)$, which is of order $C_\ell^{2}$.  An interesting feature of the trispectrum is that the index $\rho$  depends on the small-scale structure parameter $\chi$, and is therefore dependent on the mean square velocity of the string.  Observational limits are not yet strong enough to provide competitive constraints.


\subsection{Strings and 21cm radiation background}

\begin{figure}

\centering

\includegraphics[width=0.49\textwidth]{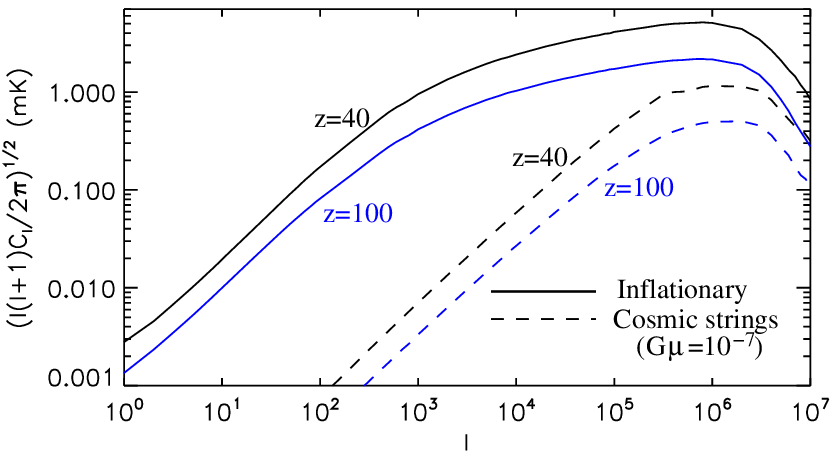}
\includegraphics[width=0.49\textwidth]{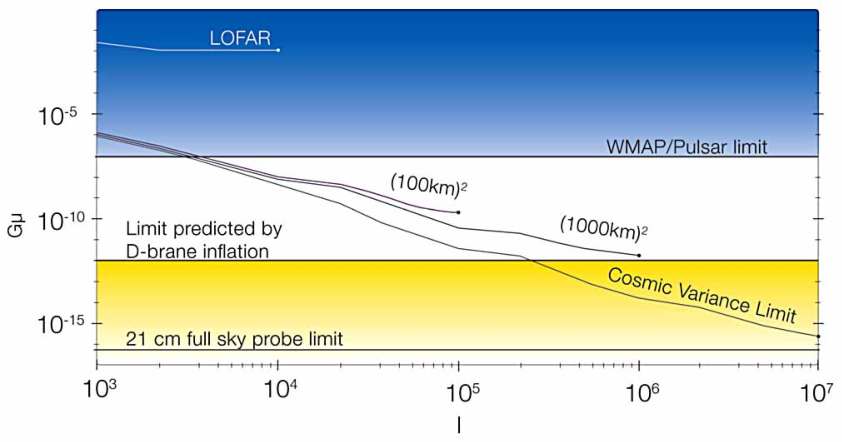}

\caption{\label{f:21cm} Left: angular power spectra from inflationary adiabatic
initial conditions (COBE normalized) and cosmic strings. Right: 
constraints from current and future experiments on
$G\mu$ for a sky fraction of 10\%, bandwidth of 0.4 MHz and integration
time of 3 years.  \cite{Khatri:2008zw}}
\end{figure}

Neutral hydrogen between us and the last scattering surface, between
 redshifts $ 30 \lesssim z \lesssim 300$, absorbs CMB radiation at 21cm
 in the local rest frame, and the measurements of the radio background
 therefore offer 
 a sensitive probe of the distribution of the matter in this range of $z$ 
\cite{ScoRee90,Loeb:2003ya,Furlanetto:2006jb,Lewis:2007kz}.
Cosmic strings contribute to the power spectrum of 21cm fluctuations \cite{Khatri:2008zw}, although at a rather low level compared to the signal from standard adiabatic perturbations (see Fig.\ \ref{f:21cm}).  The larger the collecting area the lower the $G\mu$ which can be distinguished: with over 100 km$^2$ (and a precise knowledge of the other cosmological parameters) one can reach $G\mu \sim 10^{-10}$ \cite{Khatri:2008zw}.

String-induced 21cm background perturbations are cross-correlated with the CMB \cite{Berndsen:2010xc}, but the prospects for detection even with future instruments like the Square Kilometre Array (SKA) look weak.

As in the case of the CMB, strings produce distinctive real-space features in the 21cm sky \cite{Brandenberger:2010hn}.  As they move through the neutral matter, a shocked wake of hot hydrogen accumulates behind them, leaving wedge-shaped regions of higher emission. Wakes at $z \sim 30$ from strings with  \(G\mu \simeq 6\times 10^{-7}\) should be observable with SKA \cite{Brandenberger:2010hn}. 

\section{Strings - the next generation}






We have seen that there are well-motivated inflation models which
produce cosmic strings.  There are many distinctive signals from such
models, over and above the standard inflationary ones. 
Reliable calculations require a quantitative understanding of the evolution of a network of cosmic strings from the end of inflation until today. There are two principal routes: firstly with simulations of a classical field theory such as the Abelian Higgs model, which contains all the important non-linear physics of the string width, and secondly with the classical Nambu-Goto action, with which significantly larger volumes and longer times can be simulated.  The two models give quantitatively different results for the  infinite string density and diverge considerably on the loop number density and size distribution.  

The best understood signal is the CMB, which is also the best measured.  Depending on which model one uses to simulate the string network, and which datasets are used, current observations give $2\si$ upper bounds on the string tension parameter $G\mu$ in the range $(1.6 - 6.8) \times 10^{-7}$ (see Section \ref{ss:CMBconstraints}).  Planck will improve sensitivity to $G\mu$ by a factor of 3 or so, and future B-mode missions such as CMBpol or COrE by a further factor of 3.

Strings are also well constrained by either the limits on the stochastic gravitational radiation background from millisecond pulsar timing (in the traditional Nambu-Goto scenario, Section \ref{ss:GravRad}) or by the soft $\ga$-ray background (in the case of Abelian Higgs strings, Section \ref{ss:CosmicRays}).  

An urgent task for the near future is to include gravitational radiation 
into network simulations, to check the assumption in the standard scenario that the loop size at formation is set by radiation damping.  The amplitude and spectrum of the gravitational radiation both strongly depend on this quantity and current fits to the data leave it as a free parameter, greatly reducing the constraining power. 

It is also to important to understand whether solitonic strings continue to radiate massive particles as the field theory simulations show, or whether there is some critical network length scale at which they start to behave like Nambu-Goto strings.
Nambu-Goto simulations have made impressive gains in dynamic range and the expected scaling is finally being seen loop production; however, the NG approximation assumes from the outset that massive radiation is unimportant for string networks, and so scaling behaviour in NG simulations does not in itself prove that massive radiation from solitonic strings will eventually turn off.  One way to resolve the issue would be to observe NG-like behaviour in a sufficiently large AH string simulation.  
%
However in current simulations
classical NG string networks do behave differently from classical AH ones.

Large-scale structure from strings is also an area for development.  String matter perturbations feature wakes, which are non-linear as soon as they are created, and therefore the growth of structure should be investigated using N-body simulations.

Cosmic strings are features of several models which accommodate both current particle physics and cosmological data, and there are non-trivial constraints from the combined datasets. Cosmic string phenomenology is therefore part of the wider effort to map out a theory which accounts for all data. 

It is notable that the most significant difference between elementary and solitonic strings, the small reconnection probability, was discovered as the result of a quantum calculation in string theory, which strongly motivates the drive for a more string-theoretic approach to cosmic strings.  String theory can give the decay rate of loops of string into both massless and massive states, including backreaction \cite{FStringDecay}, and therefore one can check the standard assumptions using states corresponding to realistic cosmic string configurations \cite{Skliros:2009cs}.  Calculations for networks of infinite F and D-strings remain a project for the future.





\section*{Acknowledgements}

This work was supported by the Science and Technology Research Council [grant numbers ST/G000573/1, ST/F002858/1].
I thank Richard Battye, Bjorn Garbrecht, Adam Moss, Martin Landriau, Paul Shellard, Rishi Khatri and Ben Wandelt for permission to reproduce figures from their papers, Christophe Ringeval for Fig.\ \ref{f:NGRin}, and Ken Olum, Jos\'e Blanco-Pillado and Ben Schlaer for helpful comments. Numerical calculations were performed using the UK National Cosmology Supercomputer (COSMOS, supported by SGI/Intel,
HEFCE, and STFC), and the University of Sussex HPC facility.

%

\end{document}